\documentclass{ws-ijmpa}
\usepackage[super,compress]{cite}

\begin{document}

\def\p{\partial}
\def\Lie{{\cal L}}
\def\non{\nonumber}
\def\ul{\underline}
\newcommand\mbeta{\mathring\beta}
\newcommand{\hateq}{\,\,\hat{=}\,\,}

\markboth{D. Hilditch}
{AN INTRODUCTION TO WELL-POSEDNESS AND FREE-EVOLUTION}

%
\catchline{}{}{}{}{}
%

\title{AN INTRODUCTION TO WELL-POSEDNESS AND FREE-EVOLUTION}

\author{DAVID HILDITCH}

\address{Theoretical Physics Institute, University of Jena, 07743 
Jena, Germany\\
david.hilditch@uni-jena.de}

\maketitle

\begin{abstract}
These lecture notes accompany two classes given at the NRHEP2 school.
In the first lecture I introduce the basic concepts used for analyzing 
well-posedness, that is the existence of a unique solution depending
continuously on given data, of evolution partial differential equations. 
I show how strong hyperbolicity guarantees well-posedness of the initial 
value problem. Symmetric hyperbolic systems are shown to render the initial 
boundary value problem well-posed with maximally dissipative boundary 
conditions. I discuss the Laplace-Fourier method for analyzing the initial 
boundary value problem. Finally I state how these notions extend to systems 
that are first order in time and second order in space. In the second lecture 
I discuss the effect that the gauge freedom of electromagnetism has on the 
PDE status of the initial value problem. I focus on gauge choices, 
strong-hyperbolicity and the construction of constraint preserving boundary 
conditions. I show that strongly hyperbolic pure gauges can be used to build 
strongly hyperbolic formulations. I examine which of these formulations is 
additionally symmetric hyperbolic and finally demonstrate that the system 
can be made boundary stable.
\end{abstract}

\ccode{PACS numbers: 04.20.Ex, 04.25.D-, 04.40.Nr, 41.20.-q}

\section{Well-Posedness of Evolution Partial Differential Equations}

\subsection{Introduction}	

In recent years there have been a slew of 
textbooks~\cite{BauSha10,Alc08}, review 
articles~\cite{Spe13,Pfe12,SarTig12,Hin10,Spe09,GraNov09}, and 
lecture notes~\cite{Gou10,Gou07,Pre07a} designed either as an 
introduction to numerical relativity, or as a convenient place to 
understand the state of the art of the field, not to mention classic 
texts on time evolution problems~\cite{KreLor89,GusKreOli95}. These 
resources already serve their purpose beautifully. So, happy as I was 
to be asked to teach introductory material and provide written lecture notes, 
the obvious question is; does the world need another set of introductory 
notes to hyperbolic systems? I've tried to come to a solution that presents 
the heart-and-soul of the topic as concisely as possible. In this class I 
will review concepts in the analysis of time-evolution partial differential 
equations, PDEs, proving results only sparsely. The main aim is to collect 
together, in the form of a tool-box, the necessary weapons for treating a 
given system of PDEs. I hope that where I have shamelessly copied, the 
authors of existing texts will accept my imitation as flattery. In the second 
lecture I treat electromagnetism as a model for general relativity, 
and apply each of the tools to demonstrate how they are used in practice. 
I expect that this application will be enlightening. I highlight the 
effect of gauge freedom on the PDEs analysis, for a large family of gauges. 
To my knowledge such a treatment has not appeared elsewhere, although 
free-evolution formulations of electromagnetism have of course been studied 
in the literature~\cite{KnaWalBau02,LinSchKid04,Reu04,GunGar04,GunGar05,
ReuSar05,AlcDegSal09,MacPfe13}.

In physics and applied mathematics we are frequently presented with 
systems of PDEs. Well-posedness is a fundamental property of a PDE 
problem. It is the requirement that there be a unique solution that depends 
continuously, in some norm, on given data for the problem. Without it, one has 
simply not built a reasonable mathematical abstraction of the physical problem 
at hand. The model is without predictive power, because small changes in the 
given data might result in either arbitrarily large changes in the outcome or 
that there is no solution at all. If we are given a complicated system, like the 
field equations of general relativity, we will probably have to find solutions 
numerically. But if the formulation as a PDE problem is ill-posed, 
{\it no} numerical approach can be successful! Afterall, how can one 
construct an approximation scheme that converges to the continuum solution 
if the solution doesn't exist? Therefore one might find it surprising that 
research in numerical relativity has been performed with problems that are 
ill-posed. Spontaneously on hearing that such versions of general relativity 
exist, you might think that this sounds a bit like a way of saying that the 
theory is broken, or somehow deficient. That impression is wrong. The answer is 
that for theories with gauge freedom, the precise formulation of the field 
equations as a system of PDEs affects well-posedness. And it took time for 
this fact to be recognized in the context of numerical relativity. In the second 
lecture we will carefully investigate this for electromagnetism.

The most crude way of classifying a PDE is into one of the three classes, 
Elliptic, Parabolic or Hyperbolic, names originally inspired by the 
conic-sections. The class of a PDE determines what type of data has any 
chance of producing a well-posed problem. From the intuitive point of view 
of the physicist, one might summarize their character as follows:
\begin{itemlist}
\item {\it Elliptic PDEs} have no intrinsic notion of time, and often arise as 
the steady-state, or end-state solutions of dynamical evolution, for example 
in electrostatics. The solutions to well-posed Elliptic problems are 
typically ``as smooth as the coefficients allow''. The prototype of a 
well-posed elliptic problem is the boundary value problem for the Laplace 
equation.
\item {\it Parabolic PDEs} describe diffusive processes. They have an intrinsic
notion of time, but signals travel at infinite speed. Even non-smooth 
initial data immediately become smooth as they evolve. The prototype of a 
well-posed parabolic problem is the initial value problem for the heat 
equation.
\item {\it Hyperbolic PDEs} are the best. They describe processes which are
in some sense causal; there is an intrinsic notion of time, and crucially
signals travel with {\it finite} speed. Discontinuities in initial data 
for a hyperbolic PDE will often be propagated, or may even form from 
smooth initial data. The prototype of a well-posed hyperbolic problem is 
the initial value problem for the wave equation.
\end{itemlist}
Notice that all of the prototype well-posed problems specify both a simple 
PDE, the type of data, and the domain that is appropriate. One can also 
concoct PDEs of mixed character, so this classification is certainly not 
complete. Numerical relativists have to face all three, and occasionally 
mixed classes. But in this lecture we will focus exclusively on 
hyperbolic problems and well-posedness of the initial, and initial 
boundary value problems.

\subsection{Strong hyperbolicity}	
\label{subsection:Strong}

{\it The initial value problem:} Consider a system of PDEs, which can 
be written,
\begin{align}
\p_tu=A^p\p_pu+B\,u\,,\label{eqn:System}
\end{align}
with state vector~$u$. Here I employ the summation convention, denote
$ \p_i\equiv\frac{\p}{\p x^i}$, and assume that~$i=1,2,3$. The highest 
derivative terms are called the principal part. I will sometimes refer 
to~$A^p$ as the principal matrix, although it is really a shorthand for 
three matrices, since I'm assuming that we have three spatial dimensions. 
The remaining terms on the right-hand-side of~\eqref{eqn:System} are 
called non-principal. In this lecture we will assume that the 
matrices~$A^p$ and~$B$ are constant in both time and space. We therefore 
call it a {\it linear, constant coefficient system}. The initial value, 
or Cauchy problem, is the following: specify data~$u(0,x^i)=f(x^i)$ at 
time~$t=0$, with spatial coordinates~$x^i$. What is the 
solution~$u(t,x^i)$ at later times? In other words data is 
specified {\it everywhere} in space; the domain of the solution is in 
this sense unbounded. Naturally many PDEs of interest are not linear
or do not have constant coefficients. That said, local properties of
more complicated systems are determined by the behavior of the system 
in linear approximation, which justifies the restriction.

{\it Well-posedness:} If there exist constants~$K$ and~$\alpha$, such 
that for all initial data we have the estimate,
\begin{align}
||u(t,\cdot)||\leq K e^{\alpha\,t}||f||\,,\label{eqn:WP}
\end{align}
with the~$L_2$ norm,
\begin{align}
||g||^2=\int_{\mathbf{R}^3}
g^\dagger\,g\,\textrm{d}x\,\textrm{d}y\,\textrm{d}z\,,
\end{align}
then the initial value problem for~\eqref{eqn:System} is called 
well-posed. Notice that we restrict to initial data that are 
bounded in~$L_2$.

{\it Strong hyperbolicity:} Given an arbitrary unit spatial 
vector~$s^i$, we say that the matrix 
\begin{align}
P^s&\equiv A^s=A^ps_p\,,
\end{align}
is the principal symbol of the system. The system~\eqref{eqn:System}
is called weakly hyperbolic if for every unit spatial vector~$s^i$, 
the principal symbol has real eigenvalues. If furthermore for every
unit spatial vector~$s^i$, the principal symbol has a complete set
of eigenvectors and there exists a constant~$K$, independent 
of~$s^i$, such that 
\begin{align}
|T_s|+|T_s^{-1}|\leq K\,,\label{eqn:Strong_bound}
\end{align}
where~$T_s$ is formed with the eigenvectors of~$P^s$ as columns, 
and we have the usual definition of the matrix norm~$|\cdot|$\,, the 
system is called strongly hyperbolic. If a system is strongly hyperbolic
and the multiplicity of the eigenvalues does not depend on~$s^i$ we
say that it is strongly hyperbolic of constant multiplicity. Notice 
that if the eigenvectors of the principal symbol depend continuously 
on~$s^i$ then the second condition~\eqref{eqn:Strong_bound} is 
automatically satisfied, because~$s^i$ varies over a compact set. In 
most applications we will have continuous dependence, and so checking 
for strong hyperbolicity amounts to doing a little linear algebra. 

{\it Characteristic variables:} The components of the 
vector~$v=T_s^{-1}u$ are called the characteristic variables in 
the~$s^i$ direction. Up to non-principal terms and derivatives 
transverse to the~$s^i$ direction the characteristic variables satisfy 
advection equations with speeds equal to the eigenvalues of the 
principal symbol. For this reason we will sometimes call the eigenvalues 
the speeds of the system. To see this consider,
\begin{align}
\p_tv=\Lambda_s\p_sv+(T_s^{-1}A^AT_s)\p_Av+(T_s^{-1}BT_s)v\,,
\end{align}
here~$\Lambda_s$ has the eigenvalues of the principal symbol on the 
diagonal, we denote longitudinal derivatives~$s^i\p_i=\p_s$, 
and transverse derivatives~$(\delta^i{}_j-s^js_i)\p_j$ by an upper 
case Latin index~$A$. 

{\it Well-posed~$\iff$ strongly hyperbolic:} The main result for the 
initial value problem for~\eqref{eqn:System} is that it is well-posed 
if and only if the system is strongly hyperbolic. We will need to 
Fourier transform in space, and use the convention,
\begin{align}
\tilde{f}(\omega^i)&=\frac{1}{(2\pi)^{3/2}}\int_{\mathbf{R}^3} e^{i\,\omega_jx^j}
f(x^i)\,\textrm{d}x\,\textrm{d}y\,\textrm{d}z\,.
\end{align}
The time derivative of the variables after Fourier transform is,
\begin{align}
\p_t\tilde{u}&=i\,\omega\,P^s\,\tilde{u}+B\,\tilde{u}\,,
\label{eqn:System_Fourier}
\end{align}
where we write~$\omega_i=|\omega|s_i=\omega\,s_i$. So in Fourier space 
the general solution is 
\begin{align}
\tilde{u}(t,\omega^i)&=e^{(i\,\omega\,P^s+B)t}\tilde{f}(\omega^i)\,.
\end{align}
Assume that the system is strongly hyperbolic. The key to the proof 
of well-posedness is the use of a symmetrizer. We define the Hermitian, 
positive definite symmetrizer~$H_s$ by,
\begin{align}
H_s&=T_s^{-\dagger}T_s^{-1}\,,\label{eqn:Strong_symmetrizer}
\end{align}
which satisfies the crucial property,
\begin{align}
i\,\omega\,H_sP^s+(i\,\omega\,H_sP^{s})^\dagger&=0\,.
\label{eqn:Strong_Div}
\end{align}
Note that our choice for the definition of~$H_s$ is not the most 
general that yields~\eqref{eqn:Strong_Div}; instead we could have 
chosen~$H_s=T_s^{-\dagger}C_sT_s^{-1}$, with~$C_s$ Hermitian, positive
definite, and commuting with~$\Lambda_s$. We do not require the most
general $H_s$ here, and so make do with this restriction. Define the 
norm~$||\cdot||_H$ by,
\begin{align}
||g(\cdot)||_H^2=\int_{\mathbf{R}^3}\tilde{g}^\dagger H_s\,\tilde{g}
\,\textrm{d}\omega^x\,\textrm{d}\omega^y\,\textrm{d}\omega^z\,.
\end{align}
Computing the time derivative of the norm~$||\cdot||_H$, a couple of lines 
gives the inequality,
\begin{align}
\p_t\,||u(t,\cdot)||_H^2& \leq 2\,|B|\,e^{2|B|t}\,||f(\cdot)||_H^2\,,
\end{align}
and integrating we have the well-posedness estimate, in the new 
norm~$||\cdot||_H$,
\begin{align}
||u(t,\cdot)||_H& \leq e^{|B|t}\,||f(\cdot)||_H\,.
\end{align}
But we want to obtain estimates in the $L_2$-norm. Fortunately,
with Parseval's relation~$||g(\cdot)||=||\tilde{g}(\cdot)||$, one 
can show that we have 
\begin{align}
K^{-2}\,||u(t,\cdot)||^2&\leq||u(t,\cdot)||_H^2\,,\quad&
||f||_H^2&\leq K^2||f||^2\,,
\end{align}
from which well-posedness in~$L_2$ follows. Now assume that the system
is well-posed, and consider once again the system in Fourier space. 
Suppose that at least one of the eigenvalues~$\lambda$ of the principal 
symbol is not real. Then the magnitude of the eigensolution associated 
with this eigenvalue grows like~$e^{\omega\,\textrm{Im}\lambda \,t}$, in contradiction 
with the estimate~\eqref{eqn:WP} in Fourier space. Suppose that the 
eigenvalues are real but that the principal symbol is missing one 
eigenvector of eigenvalue~$\lambda$, say, and assume that~$B=0$. The 
associated eigensolution is of the form,
\begin{align}
e^{i\,\omega\,\lambda\,t}\big[\mathbf{1}+i\,\omega\,\lambda\,t\,
(P^s-\lambda\,\mathbf{1})\big]v\,,
\end{align}
where~$v$ here is the generalized eigenvector of~$P^s$. Evidently this 
solution grows in a frequency dependent manner and, again in contradiction 
with our starting point, can not be bounded with an estimate 
like~\eqref{eqn:WP}. If there are more missing eigenvectors associated 
with~$\lambda$, the order of the polynomial in~$\omega$ increases for
the other eigensolutions. Considering~$B\ne0$ does not prevent the 
frequency dependent growth. Finally there is the possibility that 
the principal symbol is diagonalizable and has real eigenvalues but 
that the estimate~\eqref{eqn:Strong_bound} fails. As shown in 
Theorem~$2.4.1$ of~\cite{KreLor89}, this inequality is guaranteed by 
application of the Kreiss matrix theorem~\cite{Kre59}. 

{\it Variable coefficient and non-linear problems:} In applications we 
are almost never faced with linear constant coefficient problems. Given 
a linear problem of the form~\eqref{eqn:System} but now with smooth 
variable coefficients, then provided that the most general 
symmetrizer~$H_s$, described after equation~\eqref{eqn:Strong_symmetrizer}, 
can be constructed so that it is a smooth function of the 
direction~$s^i$ and the coordinates~$t,x^i$ then the well-posedness 
results carry over. So we can proceed by working in the frozen 
coefficient approximation, i.e working at an arbitrary point and 
ignoring derivatives of, or variation in the 
coefficients~\cite{SarTig12}. For non-linear problems the approach is 
to linearize at a given, possibly arbitrary, solution, and from there 
work in the frozen coefficient approximation. The price we pay is that 
well-posedness results become only local in time. To say something 
about long-term existence of solutions much more sophisticated 
methods are needed.

\subsection{Symmetric hyperbolicity}	

{\it The initial boundary value problem:} Consider now the PDE system 
similar to what we had before~\eqref{eqn:System}, 
\begin{align}
\p_tu=A^p\p_pu+F(t,x^i)\,,\label{eqn:System_Sym}
\end{align}
again with constant matrices~$A^p$, but now rather than considering 
solutions on~$\mathbb{R}^3$, let us consider trying to find solutions 
on the half-space~$x^1=x\geq0$ so that we have a boundary. We could have 
treated the non-principal forcing term~$Bu$ like this in the system we 
had before~\eqref{eqn:System}, but would not have found such a nice 
representation of the exponential growth caused by non-principal terms. 
In other words, one can think of the forcing term as being just~$F=Bu$ 
like in the previous section. We specify initial data~$u(0,x^i)=f(x^i)$ on 
the domain, and boundary conditions~$L\,u(t,x^i)\hateq g(t,x^A)$, where 
the index~$A$ here denotes that the data depends only on~$x^2=y$ and~$x^3=z$, 
with~$L$ some matrix whose form we will discuss shortly. 

{\it Strong well-posedness:} Let~$||\cdot||_\Sigma$ denote the~$L_2$
norm on the half-space, and~$||\cdot||_{\p\Sigma}$ denote the~$L_2$ norm 
in the boundary plane~$x=0$. If there exists a constant~$K_T$ for 
every~$T$, independent of the given data and forcing terms, such 
that for every~$0\leq t\leq T,$ we have the estimate,
\begin{align}
||u(t,\cdot)||_\Sigma^2+\int_0^t||u(t',\cdot)||_{\p\Sigma}^2\,\textrm{d}t'
\leq 
K_T^2\left[||f||_{\Sigma}^2+\int_0^t\Big(||F(t',\cdot)||_{\Sigma}^2
+||g(t',\cdot)||_{\p\Sigma}^2\Big)\,\textrm{d}t'
\right]\,,\label{eqn:Strong_WP}
\end{align}
then we call the problem strongly well-posed. Essentially this means 
that we can bound the solution in the bulk, and restricted to the boundary 
by the initial data, plus growth caused by either non-principal terms
or boundary data. One sometimes~\cite{GusKreOli95} sees this definition 
given without the second term on the left hand side. We will briefly 
discuss both variants below. Although these two possibilities are 
sometimes named the same way, they are distinct notions.

{\it Symmetric hyperbolicity:} If there exists a Hermitian positive 
definite symmetrizer~$H$ such that~$HA^ps_p$ is Hermitian for every 
unit spatial vector~$s^p$, then we say that the system is symmetric 
hyperbolic. Comparing the symmetrizer~$H$ with the similar 
object~$H_s$ in the proof of well-posedness of the initial value problem 
for strongly hyperbolic systems, we see the difference is that for 
symmetric hyperbolic systems the symmetrizer may not depend on~$s^p$. 
So every symmetric hyperbolic system is strongly hyperbolic, but not 
vice-versa.

{\it Maximally dissipative boundary conditions:} Since every symmetric
hyperbolic system is strongly hyperbolic, there is a matrix~$T_x$ such
that,
\begin{align}
T_x^{-1}P^xT_x=\Lambda_x=
\left(\begin{array}{cc}
\Lambda^I_x&0\\
0&\Lambda^{II}_x
\end{array}\right)\,,\label{eqn:Partition}
\end{align}
where we assume that~$\Lambda_I>0$ and~$\Lambda_{II}<0$. This last 
assumption is sometimes not met, in which case the boundary is called
characteristic in those characteristic variables with vanishing 
speed at the boundary. Characteristic boundaries complicate the analysis
considerably, in fact preventing one from showing strong well-posedness
with the type of boundary conditions that follow. I will not discuss 
such boundaries further; more information can be found 
elsewhere~\cite{SarTig12,MajOsh75,LaxPhi60,Rau85,Sec96}. Partitioning 
the characteristic variables in the x-direction~$v$ in the same way 
we write~$v=(v_I,v_{II})^\dagger$. We therefore have the condition,
\begin{align}
u^\dagger H A^x u&=v_I^\dagger H^I\Lambda^I_xv_I
+v_{II}^\dagger H^{II}\Lambda^{II}_xv_{II}\geq 
\gamma\,v_I^\dagger H^Iv_I+v_{II}^\dagger H^{II}\Lambda^{II}_xv_{II}\,,
\end{align}
for some~$\gamma>0$, and we write,
\begin{align}
T_x^\dagger\,H\,T_x&=\left(\begin{array}{cc}
H^I & 0\\
 0 & H^{II}
\end{array}\right)\,,
\end{align}
with~$\delta^{-1}\,I\leq H^I,H^{II}\leq \delta\,I$ for some~$\delta>0$. 
This block diagonal form is necessary because~$H\,A^x$ is symmetric. 
We restrict from~$L\,u\hateq g$ to consider boundary conditions of 
the form
\begin{align}
v_{II}&\hateq\kappa\,v_{I}+g\,,\label{eqn:MDBCs}
\end{align}
where~$\hateq$ denotes equality in the boundary. We assume that
\begin{align}
H^{I}\Lambda^{I}_x+\kappa^\dagger\,H^{II}\Lambda^{II}_x\,\kappa
\geq 0\,,\label{eqn:MDBC_weak}
\end{align}
which is automatically true if~$\kappa$ is sufficiently small.

{\it Strong well-posedness of symmetric hyperbolic systems with 
Maximally dissipative boundary conditions:} Consider the time derivative
of the energy~$E^2=\int_\Sigma\epsilon\,\textrm{d}V$ with~$\epsilon=u^\dagger Hu$,
which, using integration by parts, gives,
\begin{align}
\p_tE^2&=\int_{\Sigma}(u^\dagger\,H\,F+F^\dagger\,H\,u)\,\textrm{d}x\,
\textrm{d}y\,\textrm{d}z-\int_{\p\Sigma}u^\dagger HA^xu\,\textrm{d}y\,
\textrm{d}z\,,
\end{align}
if we choose boundary conditions satisfying the 
condition~\eqref{eqn:MDBC_weak}, where the the inequality does not hold 
strictly, then we have 
\begin{align}
\p_tE^2 &=\int_{\Sigma}(u^\dagger\,H\,F+F^\dagger\,H\,u)\,\textrm{d}x\,
\textrm{d}y\,\textrm{d}z-
\int_{\p\Sigma}v_I^\dagger[H^{I}\Lambda^{I}_x+\kappa^\dagger\,H^{II}\Lambda^{II}_x\,\kappa]
v_I\,\textrm{d}y\,\textrm{d}z\,,\nonumber\\
&\quad+ c_1\int_{\p\Sigma}(g^\dagger H g)\,\textrm{d}y\,\textrm{d}z\nonumber\\
&\leq\int_{\Sigma}(u^\dagger\,H\,F+F^\dagger\,H\,u)\,\textrm{d}x\,
\textrm{d}y\,\textrm{d}z 
+ c_1\int_{\p\Sigma}(g^\dagger H g)\,\textrm{d}y\,\textrm{d}z\,,
\end{align}
for some positive~$c_1$, from which the estimate~\eqref{eqn:Strong_WP} without 
the boundary term on the left hand side follows. Otherwise if the boundary 
conditions satisfy~\eqref{eqn:MDBC_weak}, but with the inequality strict, 
then playing with some inequalities, we have,
\begin{align}
&\p_tE^2+c_1\int_{\p\Sigma}(u^\dagger H u)\,\textrm{d}y\,\textrm{d}z\nonumber\\
&\qquad\qquad\leq 
\int_{\Sigma}(u^\dagger\,H\,F+F^\dagger\,H\,u)\,\textrm{d}x\,
\textrm{d}y\,\textrm{d}z
+c_2\int_{\p\Sigma}(g^\dagger H g)\,\textrm{d}y\,\textrm{d}z\,,
\end{align}
for some~$c_1,c_2>0$, from which strong well-posedness follows. 

{\it Discussion:} Using symmetric hyperbolicity to demonstrate 
well-posedness, sometimes called the energy method, is typically 
easier to approach than the method that follows, and so should be 
used whenever possible. The energy method is fantastically powerful, 
and, when it applies, can be used to estimate long-term behavior 
of solutions to variable coefficient and non-linear problems. 
Although it is not easy to construct examples of PDEs that are strongly 
but not symmetric hyperbolic in general relativity that is the very 
often the case. In which case, our only hope is the Laplace-Fourier 
method. If the PDE system is strongly, but not symmetric hyperbolic, 
then maximally dissipative boundary conditions do not guarantee 
well-posedness~\cite{CalSar03}.

\subsection{The Laplace-Fourier method}	

{\it The initial boundary value problem:} Consider once again an 
evolution system of the form~\eqref{eqn:System_Sym}\,,
\begin{align}
\p_tu&=A^p\p_pu+F(t,x^i)\,,\label{eqn:System_LF}
\end{align}
on the half-space~$x\geq0$. We assume immediately that the system is 
strongly hyperbolic with non-vanishing speeds. Characteristic 
boundaries, that is, boundaries at which the speeds vanish, can also 
be treated. In contrast to the previous case, we choose vanishing 
initial data~$u(0,x^i)=0$, but maintain inhomogeneous boundary 
conditions of the form~$L\,u(t,x^i)\hateq g(t,x^A)$ as before. We take 
the notation of the previous section. We say the the system is 
{\it strongly well-posed in the generalized sense} if there exists a 
constant~$K_T$ for every~$T$, independent of the data and forcing 
terms, such that for every~$0\leq t\leq T$, we have the estimate,
\begin{align}
&\int_0^t||u(t',\cdot)||_\Sigma^2\,\textrm{d}t'
+\int_0^t||u(t',\cdot)||_{\p\Sigma}^2\,\textrm{d}t'\leq 
K_T^2\int_0^t\Big(||F(t',\cdot)||_{\Sigma}^2
+||g(t',\cdot)||_{\p\Sigma}^2\Big)\,\textrm{d}t'\,,\label{eqn:Strong_WPgs}
\end{align}
for all boundary data~$g$. The terminology {\it in the generalized 
sense} means that we have restricted to trivial initial data, and 
that we have an estimate on the integral in time of the solution on 
the left hand side of the inequality.

{\it Laplace-Fourier Transform:} Taking the system~\eqref{eqn:System_LF} 
and Fourier transforming in the~$y$ and~$z$ directions results in a 
one-dimensional initial boundary value problem for every~$\omega^A$,
for which we need some representation of the solutions. To achieve this
we furthermore Laplace transform in time. The total transformation
is,
\begin{align} 
\hat{u}(s,x,\omega^A)&=\frac{1}{2\,\pi}\int_0^\infty
\int_{\mathbf{R}^2}e^{-s\,t+i\,\omega_Ax^A}u(t,x,x^A)\,\textrm{d}y\,\textrm{d}z
\,\textrm{d}t\,,
\end{align} 
with~$s=\eta+i\,\xi$ and~$\eta>0$. The inverse transform requires 
the contour integral, along the line~$s=\eta+i\,\xi$ with~$\eta>0$
fixed,
\begin{align} 
u(t,x^i)&=\frac{1}{(2\,\pi)^2}\oint_{-\infty}^{\infty}
\int_{\mathbf{R}^2}e^{s\,t-i\,\omega_Ax^A}\hat{u}(s,x,\omega^A)
\,\textrm{d}y\,\textrm{d}z\,\textrm{d}\xi\,,
\end{align}
which fortunately we never have to compute explicitly, because 
we have Parseval's relation, which in this context states that,
\begin{align}
\int_0^\infty\int_{\mathbf{R}^2}e^{-2\,\eta\,t}
|u(t,x,x^A)|^2\,\textrm{d}y\,\textrm{d}z\,\textrm{d}t&=
\frac{1}{2\,\pi}\int_{-\infty}^{\infty}\int_{\mathbf{R}^2}
|\hat{u}(s,x,\omega^A)|^2\,\textrm{d}\omega^y\,\textrm{d}\omega^z
\,\textrm{d}t.
\end{align}
Under this transformation we can rewrite the equations of motion as an 
ODE system
\begin{align} 
\p_x\hat{u}&=M\,\hat{u}+\hat{G}\,,\label{eqn:LF_eom}
\end{align} 
with symbol and sources,
\begin{align} 
M&=(A^x)^{-1}(s\,\mathbf{1}-i\,\omega\,A^{\hat{\omega}})\,,\quad&
\hat{G}&=(A^x)^{-1}\hat{F}\,,
\end{align} 
where we write~$\omega^A=|\omega|\,\hat{\omega}^A=\omega\,\hat{\omega}^A$,
and for later convenience define~$k=\sqrt{|s|^2+\omega^2}$, and the 
normalized frequencies~$s'=s/k$ and~$\omega'=\omega/k$.

{\it General~$L_2$ solution to the homogeneous problem:} Start by 
considering the ODE system without forcing terms~$\hat{F}$. Because 
the system is strongly hyperbolic we can assume without loss of generality 
that~$A^x$ has already been rotated to diagonal form~$\Lambda_x$. 
Assuming that the negative block of the 
partition~\eqref{eqn:Partition},~$\Lambda_x^{II}<0$ has 
dimensions~$(d\times d)$, it follows~\cite{Kre70} that~$M$ must 
have~$d$ eigenvalues with negative real part~$\kappa_i$ 
for~$i=1\dots d$. Therefore the general~$L_2$ solution 
of~\eqref{eqn:LF_eom} with vanishing~$\hat{F}$ is of the form,
\begin{align}
\hat{u}(s,x,\omega^A)&=\sum_i^d \sigma_i\,e^{\kappa_i\,x}\,
\Phi(x)\,v_i\,,
\end{align}
with~$v_i$ the eigenvector or generalized eigenvector associated 
with~$\kappa_i$, and~$\Phi(x)$ the appropriate polynomial to make 
that whole sum a sum over the eigensolutions of the ODE. 
Unfortunately strong hyperbolicity does not tell us anything 
special about the eigenvectors of~$M$, so we can not assume 
that~$M$ is diagonalizable, and therefore we must allow for this 
polynomial ansatz in the solution. The complex 
coefficients~$\sigma_i$ are to be solved for by plugging the general 
solution into the boundary conditions. 

{\it Boundary conditions and boundary stability:} As for symmetric 
hyperbolic systems, we consider here boundary conditions of the 
form~\eqref{eqn:MDBCs}, so that under Laplace-Fourier transform they 
become,
\begin{align}
\hat{u}^{II}\hateq \kappa\,\hat{u}^{I}+\hat{g}\,.
\end{align}
Where now we do not need the rotation to characteristic variables 
because we have absorbed it into the definition of~$u$. Plugging the 
general solution into the boundary conditions gives a set of linear 
equations for~$\ul{\sigma}=(\sigma_i)$,
\begin{align}
S(s,\omega)\,\ul{\sigma}=\hat{g}(s,\omega)\,,
\end{align}
for the coefficients~$\sigma_i$. If this system of equations can be 
solved such that there exists a~$\delta>0$ with,
\begin{align}
|\hat{u}(s,0,\omega^A)|\,<\,\delta\,|\hat{g}(s,\omega^A)|\,,
\end{align}
for every~$s$ and~$\omega$ with~$\eta\geq0$ then the system is called 
boundary stable. 

{\it Kreiss's symmetrizer theorem:} Boundary stability is a necessary 
condition for strong well-posedness in the generalized sense. But 
furthermore a key theorem shows that under certain 
conditions boundary stability is also sufficient. The theorem says that 
if the system~\eqref{eqn:System_LF} is either symmetric hyperbolic, or 
strongly hyperbolic of constant multiplicity, and boundary stable, 
then there exists a family of matrices that we will denote 
as~$H(s',\hat{\omega}^A)\equiv H(s',\hat{\omega})$ with smooth 
dependence on~$s'$ and~$\hat{\omega}^A$~\cite{Kre70,Agr72,Met00}, 
such that,
\begin{romanlist}[(ii).]
\item $H(s',\hat{\omega})A^x $ is Hermitian for all~$s'$, 
and~$\hat{\omega}^A$,
\item if $y$ and $h$ are vectors satisfying the boundary 
conditions~\eqref{eqn:MDBCs}, i.e~$y_{II}\hateq\kappa\,y_{I}+h$, 
then,
\begin{align}
y^\dagger\,H(s',\hat{\omega}) A^x y \geq \delta_1\,|y|^2-C\,|h|^2\,,
\end{align}
where here~$\delta_1$ and~$C$ are positive constants independent 
of~$s',\hat{\omega}^A,y$ and~$h$. 
\item There exists a constant~$\delta_2$ such that 
\begin{align}
H\,(s'\,\mathbf{1}-i\,\omega'A^{\hat{\omega}})
+(s'\,\mathbf{1}-i\,\omega'A^{\hat{\omega}})^\dagger\,H^\dagger
\geq\delta_2\,\textrm{Re}(s')\,\mathbf{1}\,.
\end{align}
\end{romanlist}
To see how this symmetrizer~$H$ can be used to show strong well-posedness 
in the generalized sense we follow the discussion of Kreiss and 
Lorenz~\cite{KreLor89}. Suppose that~$S$ is already constructed. Multiplying 
the Laplace-Fourier transformed equations of motion by~$H$ and taking the 
inner product with~$\hat{u}$ gives,
\begin{align}
-\frac{1}{k}\int_0^\infty(\hat{u}^\dagger HA^x\p_x\hat{u})\,\textrm{d}x
+\int_0^\infty\hat{u}^\dagger H\,(s'\mathbf{1}
-i\,\omega'A^{\hat{\omega}})\hat{u}\,\textrm{d}x 
=\frac{1}{k}\int_0^\infty(\hat{u}^\dagger\,H\,\hat{F})\,\textrm{d}x\,.
\label{eqn:LFTestiamte_wip}
\end{align}
We use properties~(i) and~(ii) of the theorem and obtain,
\begin{align}
2\int_0^\infty(\hat{u}^\dagger HA^x\p_x\hat{u})\,\textrm{d}x&\geq
\delta_1|\hat{u}(s,0,\omega)|^2-C|\hat{g}(s,\omega)|^2\,.
\end{align}
Taking the real part of~\eqref{eqn:LFTestiamte_wip}, multiplying 
by~$2\,k$ and using property~(iii) of~$H$ gives the estimate, 
\begin{align}
&\delta_1|\hat{u}(s,0,\omega)|^2+\delta_2\,\eta\,
\int_0^\infty|\hat{u}(s,0,\omega)|^2\textrm{d}x\nonumber\\
&\quad\quad\leq
c_1\left(\int_0^\infty|\hat{u}(s,x,\omega)|^2\textrm{d}x\right)^{1/2}
\left(\int_0^\infty|\hat{F}(s,x,\omega)|^2\textrm{d}x\right)^{1/2}
+C\,|\hat{g}(s,\omega)|^2\,.
\end{align}
with some positive~$c_1$. Inverting the Laplace-Fourier transform and 
using Parseval's relation gives strong well-posedness in the generalized 
sense. So the symmetrizer~$H$ helps even if forcing terms~$F$ are 
present. If we want to consider variable coefficient and non-linear 
problems, similar comments apply to the Laplace-Fourier method as 
those at the end of the strong hyperbolicity 
section~\ref{subsection:Strong}. 

\subsection{Second order systems}	

{\it First order in time, second order in space evolution systems:}
Very often in physics applications we are not given first order PDE 
systems like~\eqref{eqn:System}, but rather equations that are first 
order in time and second order in space, like the wave equation,
\begin{align}
\p_t\phi&=\pi\,,\quad& \p_t\pi&=\Delta\phi\,.
\end{align}
Equations of motion from a Hamiltonian fall out this way naturally. 
To analyze well-posedness of such equations we could reduce them to 
first order by introducing new variables~$d_i=\p_i\phi$ and rewriting
everything as a first order system to which the results we've been 
discussing apply. The difficulty here is that the introduction of 
these reduction variables creates constraints~$d_i-\p_i\phi=0$, and 
there is a freedom in how they can be used to make the reduction to 
first order. Fortunately it is not necessary to take care of these 
subtleties, because conditions under which~``good'' reductions exist 
have been analyzed in the literature~\cite{GunGar05}. Recently these 
calculations have been extended to treat high order systems of hyperbolic 
equations~\cite{HilRic13a}. So consider the second order in space evolution 
system,
\begin{align}
\p_tv&=A_1^i\p_iv+A_1v+A_2w+F_v\,,\nonumber\\
\p_tw&=B_1^{ij}\p_i\p_jv+B_1^i\p_iv+B_2^i\p_iw+B_2w+F_w\,,\label{eqn:FT2S}
\end{align}
as in the first order case assume that the coefficient matrices are 
constant. We call the matrix 
\begin{align}
A^p{}_i{}^j&=\left(\begin{array}{cc}
A_1^j\delta^p{}_i  & A_2\delta^p{}_i\\
B_1^{pj} & B^p_2 
\end{array}\right)\,,
\end{align}
the principal part matrix of the system. The indices~$i,j,p$ run 
over all spatial directions. The index~$i$ labels blocks of rows, 
and~$j$ blocks of columns. 

{\it Strong hyperbolicity and characteristic variables:}  Given an 
arbitrary unit spatial vector~$s^i$, we call 
the matrix,
\begin{align}
P^s&=S^iA^p{}_i{}^jS_js_p=
\left(\begin{array}{cc}
A_1^js_j & A_2\\
B_1^{pj}s_ps_j & B_2^ps_p
\end{array}\right).
\end{align}
with the abbreviation
\begin{align}
S_i&=\left(\begin{array}{cc}
s_i&0\\
0&\mathbf{1}
\end{array}\right)\,,
\end{align}
the principal symbol of the system. If for every unit spatial vector the 
eigenvalues of the principal symbol are real, we call the system weakly 
hyperbolic. If furthermore for every unit spatial vector~$s^i$ the 
principal symbol has a complete set of eigenvectors such that there
exists a constant~$K$ independent of~$s^i$, such that 
\begin{align}
|T_s|+|T_s^{-1}|\leq K\,,
\end{align}
where~$T_s$ is formed with the eigenvectors of~$P^s$ as columns, the 
system is called strongly hyperbolic. This condition is equivalent to 
well-posedness of the initial value problem, where now the norm contains 
first spatial derivatives of~$v$, $\p_iv$. It is also equivalent, at 
least in three spatial dimensions, to the existence of a strongly 
hyperbolic first order reduction~\cite{GunGar05} and likewise a strongly
hyperbolic pseudo-differential reduction~\cite{Tay81,NagOrtReu04}. I do not 
know of a place where this equivalence has been shown in higher spatial dimensions
in physical space, but the result holds for pseudo-differential 
reductions in arbitrary spatial dimensions. The characteristic variables 
of the second order in space system are defined to be the components of,
\begin{align}
u=T_s^{-1}\left(\begin{array}{c}
\p_sv\\
w\end{array}\right)\,.
\end{align}
Strong hyperbolicity implies the existence of a complete set of 
characteristic variables just like in the first order case.

{\it Symmetric hyperbolicity:} We call a symmetric matrix~$H^{ij}$,
independent of~$s^i$, such that 
\begin{align}
S_iH^{ij}A^p{}_j{}^ks_pS_k=(S_iH^{ij}A^p{}_j{}^ks_pS_k)^\dagger.
\label{eqn:FT2S_sym}
\end{align}
for every spatial vector~$s^i$, a candidate symmetrizer. A positive
definite candidate symmetrizer is called a symmetrizer. A system 
with a symmetrizer is called symmetric hyperbolic. The symmetrizer
can be used to define a conserved energy, at least up to non-principal
terms. Slightly abusing notation, the energy density is, 
\begin{align}
\epsilon&=(\p_iv,w)\,H^{ij}(\p_jv,w)^\dagger\nonumber\\
&=\p_iv^\dagger\,H_{vv}^{ij}\p_jv
+\p_iv^\dagger\,H_{vw}^{i}w
+w\,H_{vw}^{j\,\dagger}\p_jv^\dagger
+w^\dagger\,H_{ww}w\,.
\end{align}
In simple cases, say for the wave equation, this ``PDEs energy'' may 
correspond to a true physical energy~\cite{HilRic10,RicHil11}. In general a
Hamiltonian for the system~\eqref{eqn:FT2S} 
guarantees a candidate symmetrizer, but {\it not} a symmetrizer. This 
definition of symmetric hyperbolicity for second order in space systems 
is equivalent to the existence of a first order reduction 
of~\eqref{eqn:FT2S} that is symmetric hyperbolic according to the 
definition for first order systems. Maximally dissipative boundary conditions can be 
defined for second order in space systems in a similar way to first 
order systems, and can again be used to guarantee estimates of the 
solution including boundary data. 

{\it The Laplace-Fourier method:} The Laplace-Fourier method applies 
to the second order in space system\eqref{eqn:FT2S} 
straightforwardly~\cite{KreWin06,KreOrtPet10,SarTig12}. For brevity 
let us assume that there are as many~$v$'s as~$w$'s, and that~$A_2$ is 
invertible. Under this assumption, and that of trivial initial data, 
grouping all of the non-principal terms together, we can Laplace-Fourier 
transform, and arrive at,
\begin{align}
s^2\,\hat{v}&=
A^{xx}\p_x\p_x\hat{v}+2\,i\,\omega\,A^{x\hat{\omega}}\p_x\hat{v}
-\omega^2\,A^{\hat{\omega}\hat{\omega}}\,\hat{v}
+s\,B^x\p_x\hat{v}+i\,\omega\,B^{\hat{\omega}}\,\hat{v}+\hat{F}\,,
\end{align}
with the shorthands
\begin{align}
A^{ij}&=A_2B_1^{ij}-A_2B_2^{(i}A_2^{-1}A_1^{j)}\quad& B^i&=
A_1^i+A_2B_2^iA_2^{-1}\,,
\end{align}
with~$A^{ij}$ symmetric in~$i$ and~$j$. Assuming that the system is 
strongly hyperbolic with non-vanishing speeds, we can introduce the 
reduction variables~$D\hat{v}=k^{-1}\,\p_x\hat{v}$, and manipulate 
the second order ODE system to end up with 
\begin{align}
\p_x\hat{u}&=M\,\hat{u}+\hat{G}\,,
\end{align}
with the symbol,
\begin{align}
M(s,\omega^A)&=k \left(\begin{array}{cc}
0& \mathbf{1}\\
A&B
\end{array}\right)\,.
\end{align}
with the lower two blocks given by,
\begin{align}
A&=(A^{xx})^{-1}[s'^2\,\mathbf{1}+\omega'^2\,A^{\hat{\omega}\hat{\omega}}]\,,
\quad& B&=(A^{xx})^{-1}[s'\,B^x+2\,i\,\omega'\,A^{x\hat{\omega}}]\,.
\end{align}
This type reduction is called a pseudo-differential reduction to first 
order. We can construct the general solution to the ODE and consider 
boundary stability as before. Ultimately the norms that we use to 
estimate solutions will again contain first spatial derivatives 
of~$v$.

\subsection{Summary}

\begin{figure}[pt]
\centerline{\psfig{file=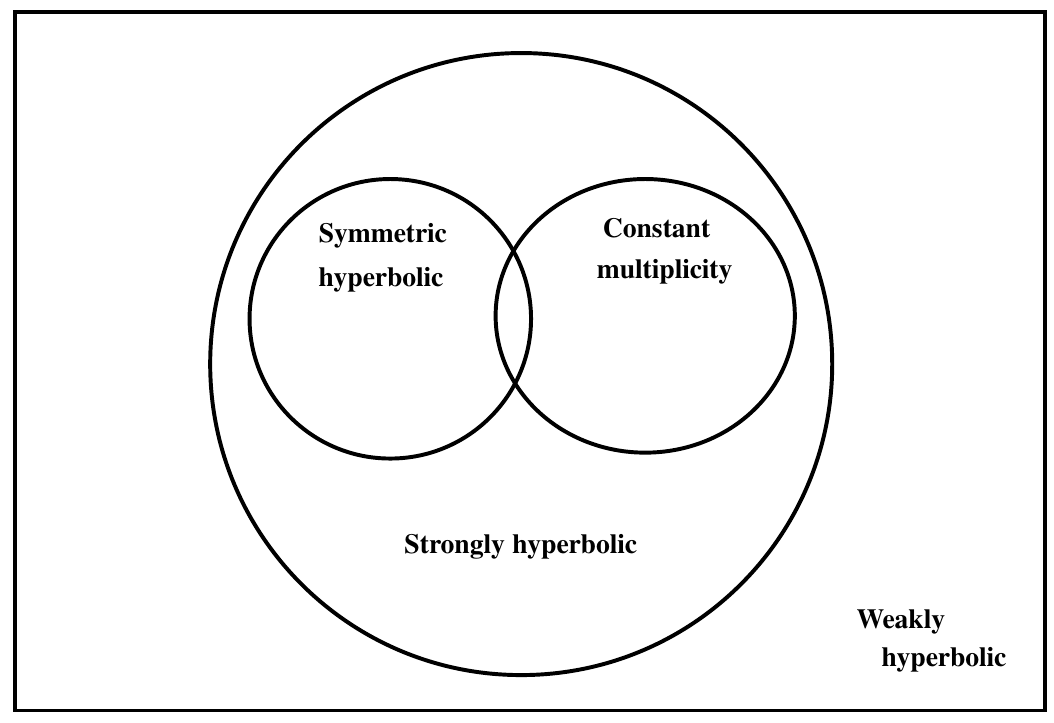,width=9cm}}
\vspace*{8pt}
\caption{A schematic summary of the different levels of hyperbolicity.
Every strongly hyperbolic system is weakly hyperbolic. Every symmetric 
hyperbolic system is strongly hyperbolic, and furthermore there is an
intersection between systems that are strongly hyperbolic of constant
multiplicity and symmetric hyperbolic, although the notions are not
coincident.\label{fig:Hyp_Venn}}
\end{figure}

I have given the definitions of well-posedness and hyperbolicity for 
initial and initial boundary value problems for first order systems. 
The relationship between the various definitions are summarized in 
Fig.~\ref{fig:Hyp_Venn}. I sketched how these definitions are extended 
to first order in time, second order in space systems. Obviously the 
description here is superficial, and I therefore recommend that you take a 
look at the books and review articles cited throughout. In the next 
class we will see how these methods are applied to electromagnetism as 
a free-evolution system. From the point of view of hyperbolicity analysis
electromagnetism is a very satisfactory model for general relativity
since it has both constraints and gauge freedom.

\section{Free Evolution Formulations of Electromagnetism}

\subsection{Introduction}	

In the first lecture we saw several notions of hyperbolicity, and 
that they are useful in different contexts. The moral of the story 
was three-fold:

\begin{romanlist}[(ii).]
\item Strong hyperbolicity is good enough for the initial 
value problem, and is easy to check-- there is no good excuse not 
to bother!

\item Symmetric hyperbolicity, or the energy method, is 
good for the IBVP, and is the 
preferred approach whenever it applies because it is 
simple to apply.
\item The Laplace-Fourier method can be used to analyze 
well-posedness of the IBVP for PDEs that are strongly 
hyperbolic of constant multiplicity. If the system is 
only strongly hyperbolic then more work is needed to 
make definite statements because the theory is not 
complete. The Laplace-Fourier method applies to a larger 
class of boundary conditions than those that can be 
treated with the energy method, but the algebraic 
manipulation required is usually more involved.
\end{romanlist}
In this class we will apply these notions to Maxwell's 
theory of electromagnetism, and will see that new 
complications arise. I want you to think of Maxwell as a 
model for general relativity; qualitatively nearly all of 
the same features are present. But the lower valence of 
the tensor fields make the whole thing easier to treat 
in a short time. I have prepared a number of mathematica 
notebooks to accompany the lecture, so that you can see 
how all of the various steps can be made in practice. The 
qualitative difference between the Maxwell equations and 
those of the previous lecture is gauge freedom. We therefore 
have to work a little before we can apply those notions. 
Dirac's theory of constrained Hamiltonian systems~\cite{Dir64} tells 
us that there is a relationship between the gauge freedom 
and the constraints of the theory, but we won't use 
that deeply here. Instead I want to go through the various 
features of Maxwell, but impress upon you now that the 
structure we will discover in the equations of motion 
falls out because of the Hamiltonian form of the theory.
In Helvi Witek's lectures~\cite{Wit13} one popular formulation, called 
BSSNOK~\cite{BauSha98,ShiNak95,NakOohKoj87} is described in a form 
quite similar to what we will have for electromagnetism, but with 
dimension as a parameter.

\subsection{The vacuum Maxwell equations}

\subsubsection{Hamiltonian and equations of motion} 

Let us start with the Hamiltonian for electromagetism, in curved 
space, which is given by
\begin{align}
H&=\int_\Omega\tfrac{1}{2}\big[(D\times A)_i(D\times A)^i + E_iE^i\big]
-\Phi\,D^iE_i\,\textrm{d}V\,.\label{eqn:Hamiltonian}
\end{align}
We have canonical positions~$A_i$, and 
momenta~$\tilde{\pi}^i=-\sqrt{\gamma}\,E^i$, and define the curl by
\begin{align}
(D\times A)^i=\epsilon^{ijk}D_jA_k\,,
\end{align}
with~$\epsilon^{ijk}$ the Levi-Civita tensor. From this and Hamilton's 
equations, we obtain the equations of motion 
\begin{align}
\p_tA_i&=-\alpha\,E_i-D_i\Phi+\Lie_\beta A_i\,,\nonumber\\
\label{eqn:AEdot}
\p_tE^i&=\big(D\times\alpha\,[D\times A]\big)^i
+\alpha K E^i+\Lie_\beta E^i\,,
\end{align}
where for personal preference I work with the electric field~$E_i$ 
rather than the canonical momentum~$\tilde{\pi}^i$. Here I am using the 
standard notation for the lapse function~$\alpha$, the shift 
vector~$\beta^i$, spatial metric~$\gamma_{ij}$, extrinsic curvature~$K_{ij}$
and the volume element~$\textrm{d}V$. I write the covariant derivative 
compatible with~$\gamma_{ij}$ as~$D_i$. The lapse measures elapsed 
proper time between neighboring time slices. The shift describes how 
spatial coordinates are translated across times slices, and the 
extrinsic curvature is just the second fundamental form of the spatial 
slice as embedded in the spacetime. If you've not met these quantities 
before, you can just ignore them by setting,
\begin{align}
\alpha&=1,\quad& \beta^i&=0,\quad& \gamma_{ij}&=\delta_{ij},\quad& K_{ij}&=0,
\end{align}
recovering at least a subset of the equations that you're used to 
from electrodynamics~\cite{Jac99}. The~$3+1$ decomposition of 
spacetime is described beautifully elsewhere~\cite{Gou07}. Note that,
as in other sources~\cite{AlcDegSal09} we could choose to work with 
the magnetic field~$B^i=(D\times A)^i$, but will not do so, because 
we do not want to lose the analogy between relativity and 
electromagnetism. Formulations similar to that which we consider 
here have also been examined~\cite{KnaWalBau02,GunGar05} in the 
literature.

\subsubsection{Constraints}

Of course we have the constraint that the divergence of the electric 
field vanishes,
\begin{align}
M=-D_i\tilde{\pi}^i=D_iE^i=0,
\end{align}
which is obtained from the Hamiltonian~\eqref{eqn:Hamiltonian} by varying 
with respect to the Lagrange multiplier~$\Phi$, which we will call 
the gauge field. We call this restriction the momentum constraint.
Computing the time derivative using the equations of 
motion~\eqref{eqn:AEdot} of the momentum constraint reveals
\begin{align}
\p_tM&=\alpha KM+\Lie_\beta M.
\label{eqn:HamMdot}
\end{align}
Notice that the right hand contains only terms multiplied 
by the constraint, or its spatial derivative. So if we start 
with constraint satisfying initial data, then the constraints 
will remain satisfied in the time development of the data. This fact 
is expressed in different ways in the literature. It is sometimes said 
that the constraints propagate. Anyway, for numerical applications this means 
that all we need do is specify initial data satisfying the constraint, 
and then integrate up the equations of motion in time. This approach 
is called {\it free-evolution}. It is the main approach for treating 
the field equations of general relativity numerically. This 
is how we arrive at a evolution equations that we can treat 
with the methods of the last lecture. A few comments are in 
order:

\begin{romanlist}[(ii).]
\item In applications using the free-evolution approach, 
constraint violation is inevitable because of numerical error. 
The best we can hope for is that as we throw more resolution at the 
problem we can make the errors arbitrarily small, in which case the 
numerical analyst might consider it solved. If sufficient 
computational power for a desired error is {\it actually at 
their disposal} the computational physicist will also consider 
it solved. 

\item In the free-evolution approach one must analyze the 
PDE properties of the system without assuming that the 
constraints are satisfied. This is exactly because due to 
numerical error we are really computing in the larger phase 
space which includes violations. For an introduction to solving 
the constraints of general relativity, see Hirotada Okawa's lectures 
later in the school~\cite{Oka13}. It is sometimes hard, or we are too 
lazy, to solve the constraints in the initial data. Although this is 
in principle wrong, if one can be confident that the violation is not 
the leading source of error in the numerical calculation, it may be 
well justified. 

\item A rather different approach is to resolve the 
constraints after every time-step, which is called 
constrained-evolution. With this method one expects to arrive 
at a mixed elliptic-hyperbolic problem, which can not 
be treated with the methods discussed in the last lecture.
\end{romanlist}

\subsubsection{Gauge freedom and the pure gauge system}

The equations of motion~\eqref{eqn:AEdot} are 
invariant under the transformation, 
\begin{align}
A_i\to A_i-D_i\psi\,,
\end{align}
with some arbitrary scalar function~$\psi$. Suppose that 
we are given constraint satisfying initial data. What is 
the difference of in the time development with and 
without applying a gauge transformation to the initial 
data? It is easy to see that the pure gauge 
field~$\psi$ evolves according to
\begin{align}
\p_t\psi&=\Delta[\Phi]+\Lie_\beta\psi\,,
\label{eqn:PG_Psi}
\end{align}
where~$\Delta[\Phi]$ denotes the difference in $\Phi$
induced by the gauge change in the initial data.

It still wouldn't quite be proper to start our well-posedness 
analysis, because the Hamiltonian does not determine the 
field~$\Phi$. And we can't very well analyze a set of 
equations if we don't know what they are. So for this we 
need a gauge choice, by which we mean a choice of~$\Phi$. 
We will consider an equation of motion of the form 
\begin{align}
\p_t\Phi&=-\mu\,\alpha^2\p^iA_i+\Lie_\beta\Phi.
\label{eqn:Gauge}
\end{align}
for~$\Phi$, with~$\mu$ a scalar field that does not 
depend on any of the Maxwell fields. I'm making this restriction 
so that the field equations remain linear. 
We write~$\p^iA_i=\gamma^{ij}\p_iA_j$. Note that 
choosing the partial derivative in the divergence here {\it is} 
intentional. If we took the covariant derivative then once we 
considered the system coupled to general relativity the resulting terms 
would not be minimally coupled. In principle we could just choose the 
gauge field as an apriori function, or indeed require that it satisfy an 
elliptic equation. But we will restrict ourselves exclusively 
to evolution gauges of the form~\eqref{eqn:Gauge}.

With this choice, how will~$\Delta[\Phi]$, 
the induced change from a gauge transformation in the initial 
data, evolve in time? Computing the time derivative of the difference 
of the two~$\Phi$'s reveals
\begin{align}
\p_t\Delta[\Phi]&=-\mu\alpha^2
\p^i\p_i\psi+\Lie_\beta\Delta[\Phi]\,.
\label{eqn:PG_Phi}
\end{align}
Interestingly we have to arrived at a closed 
subsystem~\eqref{eqn:PG_Psi},\eqref{eqn:PG_Phi} for the 
evolution of the change in gauge, which we call the 
pure gauge system. The obvious question is now: what is the 
relationship, if any, between hyperbolicity of this pure 
gauge system and the Maxwell equations under the 
choice~\eqref{eqn:Gauge}? We are nearly ready to try and 
tackle that problem, but one complication remains to be 
dealt with first. 

\subsubsection{Expanded phase space}

The next step in the construction of our free-evolution formulation 
is to expand the phase space with another variable, which we 
call~$Z$. The new variable is constrained to vanish on physical 
solutions. It needs an equation of motion, for which we choose
\begin{align}
\p_tZ&=\alpha D^iE_i-\alpha\,\kappa\,Z
+\Lie_\beta Z\,,\label{eqn:Zdot}
\end{align}
with~$\kappa$ some constant parametrizing the 
{\it constraint damping}~\cite{WeyBerHil11,Owe07,GunGarCal05,BroFriHub99}, 
which I won't have time to discuss further. So you can see that if both 
the constraints~$Z$ and the~$M$ are initially satisfied then~$Z$ will 
stay satisfied, provided that the inclusion of the new constraint does 
not break the momentum constraint. If we wanted to, here we could 
have chosen any amount of the momentum constraint in the 
right hand side of this equation of motion. Instead I have 
made a choice, the first term, that will turn out to be convenient 
later.

Given that we want to find solutions of the Maxwell 
equations, this initially seems a bit strange. Why would we 
want to expand the the solution space with {\it more} freedom 
to be wrong? The answer is that the PDE properties of the 
problem we want to solve are affected by this expansion, and 
typically favorably. It is imperative that we solve 
well-posed problems, so we are forced to consider the 
expansion. If we were to insist on the whole set of 
equations of motion, including those for the gauge 
choice~\eqref{eqn:Gauge} as coming from a Hamiltonian, you 
can think, roughly speaking, of the new constraint~$Z$ as 
the canonical momentum of the gauge field~$\Phi$. In text 
books you normally see that type of construction only for 
the Lorenz gauge~$\mu=1$, often in the context of 
quantum electrodynamics. The equivalent type of construction 
can be made for general relativity~\cite{Bro10,BonBonPal10,HilRic10,Bro08}, 
but from the point of view of the physical system there is no reason to 
restrict the equations of motion in this way. Only physical 
quantities need satisfy Hamilton's equations.

\subsubsection{Fully expanded equations of motion}

The last step is to choose how the new constraint is to 
be coupled. For this we are free to make parametrized 
additions of~$Z$ or its derivative to the other equations 
of motion. As in~\eqref{eqn:Zdot}, I will make a convenient
choice now. The full equations of motion are then taken to be
\begin{align}
\p_tA_i&=-\alpha E_i-D_i\Phi+\Lie_\beta A_i\,,\nonumber\\
\p_tE^i&=\big(D\times\alpha\,[D\times A]\big)^i
+\alpha D^iZ
+\alpha K E^i+\Lie_\beta E^i\,,\nonumber\\
\p_t\Phi&=-\mu\,\alpha^2[\p^iA_i+Z]
+\Lie_\beta\Phi\,,\nonumber\\
\p_tZ&=\alpha\,D_iE^i-\alpha\,\kappa\,Z+\Lie_\beta Z\,.\label{eqn:Max_Full}
\end{align}
You should think about the consequences of making different 
choices after we see how the well-posedness analysis of the 
next section. The system is subject to constraints 
\begin{align}
Z&=0\,,\quad& M&=D^iE_i=0\,.
\end{align}
The constraint subsystem is still closed. Computing the 
time derivative of the momentum constraint, which is altered 
from~\eqref{eqn:HamMdot}, reveals 
\begin{align}
\p_tM&=\alpha D^iD_iZ+D^i\alpha\,D_iZ+\alpha KM
+\Lie_\beta M\,.
\label{eqn:ExpandMdot}
\end{align}
As we hoped when introducing the new constraint, the 
momentum constraint is not broken by~$Z$. Assuming uniqueness of solutions
to the subsystem, if the constraints are initially satisfied then they will 
remain so as the solution develops in time, so a free-evolution approach is 
justified. As in the case of the pure gauge system, one might wonder 
how, if at all, hyperbolicity of the closed constraint 
subsystem is inherited by the full equations of motion. 
We will consider these questions in the next section.
Finally we have arrived at the formulation of the Maxwell 
equations that we will analyze with the methods of the last 
lecture. The obvious difference between our version of the Maxwell 
equations is that in general the given background lapse, shift, spatial 
metric and extrinsic curvature are constant in space and time. As 
discussed in the first class, we will side-track the issue by working 
in a neighborhood of an arbitrary point so that to a good approximation 
these coefficients are constant, and so we can ignore them. This is 
called the frozen coefficient approximation. Comparing 
with~\eqref{eqn:FT2S}, we see that the~``$v$'' variables are $A_i$ 
and~$\Phi$, whilst the~``$w$'' variables are~$E_i$ and~$Z$. Notice that 
if we want to keep the shape of~\eqref{eqn:FT2S}, we are not free to add 
derivatives of the momentum constraint to~$E_i$'s equation of motion. If we 
were to do so, we would have to start thinking in terms of a mixed 
hyperbolic-parabolic system.

\subsection{Well-posedness analysis}

\subsubsection{Strong hyperbolicity}

We work in the frozen coefficient approximation and discard 
non-principal terms. We need to consider the principal symbol for an 
arbitrary unit vector~$s^i$. To make the analysis tidier we use this 
unit vector to $2+1$~decompose the vector quantities of electromagnetism, 
writing,
\begin{align}
\p_sA_i&=s_i[\p_s^2\psi] - s_iZ
+\perp^A_i[\p_sA_A]\,,
\quad& E_i&=s_iE_s+\perp^A_iE_A\,,
\end{align}
where we have defined the projection operator
\begin{align}
\perp^i_j=\delta^i{}_j-s^is_j\,,
\end{align}
and use upper case Latin indices~$A,B,C$ to denote 
projected objects. In the frozen coefficient approximation 
all derivatives of background quantities, including~$s^i$ 
vanish. Choosing to use the new variable~$[\p_s^2\psi]=\p_sA_s+Z$ may 
not initially seem natural, but the reason, which you can easily 
guess from the name, will rapidly become clear. The principal symbol 
splits into three decomposed blocks that can be read off from 
\begin{align}
\p_t[\p_s^2\psi]&\simeq
-\p_s[\p_s\Phi]
+\beta^s\p_s[\p_s^2\psi]\,,\quad&
\p_t[\p_s\Phi]&\simeq-\mu\,\alpha^2
\p_s[\p_s^2\psi]
+\beta^s\p_s[\p_s\Phi]\,,
\label{eqn:PG_coupled}
\end{align}
should naturally be compared with the principal symbol 
of the pure gauge system. The remaining ``scalar'' equations are,
\begin{align}
\p_tZ&\simeq\alpha\,\p_sE_s+\beta^s\p_sZ\,,\quad&
\p_tE_s&\simeq\alpha\,\p_sZ+\beta^s\p_sE_s\,,
\label{eqn:Cons_coupled}
\end{align}
which should be compared with the principal symbol of the 
constraint subsystem, and finally,
\begin{align}
\p_t[\p_sA_A]&\simeq-\alpha\,
\p_sE_A+\beta^s\p_s[\p_sA_A]\,,\quad&
\p_tE_A&\simeq-\alpha\,
\p_s[\p_sA_A]
+\beta^s\p_sE_A\,,\label{eqn:Physical}
\end{align}
which is decoupled from both the ``gauge'' and ``constraint''
variables of the first two blocks. So very happily we find that both 
the the pure gauge and constraint principal symbols, which 
can be read off from the pairs~\eqref{eqn:PG_Psi},\eqref{eqn:PG_Phi} 
and~\eqref{eqn:Zdot},\eqref{eqn:ExpandMdot} are inherited by 
the formulation. This is not a coincidence and can be shown 
for a large class of constrained Hamiltonian systems~\cite{HilRic13}. 
These type of questions have also been studied for systems with 
constraints without requiring gauge freedom~\cite{Reu04,GunGar04}.
If we had chosen to add the constraints to the evolution equations 
differently, we could have ended up with the ``constraint'' variables 
in the right hand sides of~\eqref{eqn:PG_coupled}. But you should 
convince yourself that we could not have a formulation with the 
``gauge'' variables in the right hand sides of 
either~\eqref{eqn:Cons_coupled} or~\eqref{eqn:Physical}. The principal 
symbol of each block is,
\begin{align}
P_\mathcal{G}^s&=
\left(
\begin{array}{cc}
\beta^s&-1\\
-\mu\alpha^2&\beta^s
\end{array}\right)\,,\quad&
P_\mathcal{C}^s&=
\left(\begin{array}{cc}
\beta^s&\alpha\\
\alpha&\beta^s
\end{array}\right)\,,\quad&
P_\mathcal{P}^s&=
\left(
\begin{array}{cc}
\beta^s&-\alpha\\
-\alpha&\beta^s
\end{array}\right)\,,
\end{align}
respectively. The pure gauge block has 
eigenvalues~$\lambda_{\pm\mu}=\beta^s\pm\sqrt{\mu}\alpha$. We 
assume that~$\alpha>0$, so the only requirement for weak hyperbolicity 
is that~$\mu\geq0$. With this restriction, the block is 
diagonalizable if~$\mu>0$. The other two blocks both 
have eigenvalues~$\lambda_{\pm}=\beta^s\pm\alpha$, corresponding to the 
speed of light in the~$s^i$ direction, and are furthermore 
diagonalizable. The eigenvectors of each block are
\begin{align}
&\big(\pm1,\,\sqrt{\mu}\alpha\big)^T\,,\quad&
&\big(\pm1,\,1)^T\,,\quad& &\big(\pm1,\,1)^T\,,
\end{align}
respectively, and have characteristic variables  
\begin{align}
&[\p_s\Phi] \pm [\p_s^2\psi]\,,\quad&
 &E_s\pm Z\,,\quad&
 &E_A \pm [\p_sA_A]\,,
\end{align}
with speeds~$\lambda_{\pm\mu},\lambda_{\pm},$ and~$\lambda_{\pm}$. 
These calculations are performed in the mathematica 
notebook~{\tt Maxwell\_Strong.nb} which accompanies the 
lecture.

\subsubsection{Symmetric hyperbolicity}

In this section we will try to see which of family of 
the gauge conditions~\eqref{eqn:Gauge} result in a PDE 
system that is symmetric hyperbolic. The answer is 
``every gauge that is strongly hyperbolic''. Therefore 
before we start I want to give a word of warning; this 
result is a special feature of these formulations of the Maxwell 
equations, and does not necessarily carry over to other theories that 
we are interested in. In particular it is not true for relativity, 
as can be seen for popular gauge choices in numerical 
relativity~\cite{HilRic10}. More generally, suppose that 
we have a gauge choice for which the pure gauge subsystem 
is either strongly or symmetric hyperbolic. It has not been shown that 
there is necessarily a free-evolution formulation that is symmetric 
hyperbolic with that gauge. One might take the view that strongly, but 
not symmetric hyperbolic formulations are objectively worse than those 
that are symmetric hyperbolic. But sometimes in applications the 
choice that ``works'' might be the mathematically weaker 
one; so we can not always just choose the symmetric hyperbolic 
system. This can be problematic because, as we have seen, establishing
well-posedness of the initial boundary problem is tricky for  
generic systems that are only strongly hyperbolic, because there may be 
no applicable theory. With a bigger gauge 
freedom, it could even be that the choice of pure gauge that is useful is 
not symmetric hyperbolic, and then of course the expectation is that we 
can't use that gauge to build a strongly hyperbolic formulation. 
These are the types of problems that can happen for the Einstein 
equations.

So that I can work with the same set of equations in this 
section and the next, I start by writing the principal part 
of the system~\eqref{eqn:Max_Full} in fully second order form,
\begin{align}
\p_0^2\Phi&\simeq \mu\gamma^{ij}
\p_i\p_j\Phi\,,\quad&
\p_0^2A_i&\simeq\gamma^{jk}\p_j\p_kA_i
+\big(\tfrac{1}{\mu}-1\big)\p_0\p_i\Phi\,.
\end{align}
where we have defined~$\p_0=(\p_t-\beta^i\p_i)/\alpha$. Notice 
that if we take the Lorenz gauge, $\mu=1$, each variable satisfies a 
decoupled wave equation. This change of variables does not affect 
the PDE properties of the system. Why not? From this one writes the 
principal part matrix as 
\begin{align}
A^p{}_{i\,k}{}^{j\,l}&=\left(\begin{array}{cccc}
0 & 0 & \delta^p{}_i & 0 \\
0 & 0 & 0 & \delta^p{}_i\delta^l{}_k\\
\mu\,\gamma^{pj} & 0 & 0 & 0\\
0 & \gamma^{pj}\delta^l{}_k
  & \big(\frac{1}{\mu}-1\big)\delta^p{}_k & 0
\end{array}\right)\,,
\end{align}
where here we've picked up indices from the fields in the principal 
part matrix, but this is completely compatible with the way we defined 
the principal part matrix for second order in space systems. We make 
an ansatz for an energy density~$\epsilon$ with
\begin{align}
\epsilon&=u_{jm}^\dagger H^{ij\,km}u_{ik}\,,\quad& 
u_{ik}=(\p_i\Phi\,,\p_iA_k\,,\p_0\Phi\,,\p_0A_k)^\dagger\,,
\end{align}
and a parametrized ansatz for~$H^{ij\,km}$,
\begin{align}
H^{ij\,km}&=\left(\begin{array}{cccc}
h_{11}^1\gamma^{ij} & 0 & 0 & h_{14}^1\gamma^{ik} \\
0 & h_{22}^1\gamma^{ij}\gamma^{km}+2\,h_{22}^2\gamma^{k(i}\gamma^{j)m} 
+ 2\,a_{22}^1\gamma^{k[i}\gamma^{j]m}
& h_{23}^1\,\gamma^{im} & 0\\
0 & h_{23}^1\,\gamma^{jk} & h_{33}^1 & 0 \\
h_{14}^1\gamma^{jm} & 0 & 0 & h_{44}^1\gamma^{km}
\end{array}\right)\,.
\end{align}
Imposing conservation of the energy, or in other words 
Hermiticity of,
\begin{align}
&S_i\,H^{ij\,mn}\,A^p{}_{j\,m}{}^{k\,l}\,s_p\,S_k=\\
&\left(\begin{array}{cccc}
0 & h_{14}^1s^l & h_{11}^1+h_{14}^1(\frac{1}{\mu}-1) & 0 \\
h_{23}^1\,\mu\,s^n & 0 & 0 & h_{22}^1\gamma^{ln}+2h_{22}^2s^ls^n\\
h_{33}^1\,\mu & 0 & 0 & h_{23}^1\,s^l \\
0 & h_{44}^1\,\gamma^{ln} & 
\frac{1}{\mu}\big(h_{44}^1(1-\mu)+h_{14}^1\,\mu\big)s^n & 0
\end{array}\right)\,,\nonumber
\end{align}
for every spatial vector~$s^i$, see the last 
lecture~\eqref{eqn:FT2S_sym}, gives the conditions,
\begin{align}
h_{14}^1&=h_{23}^1\,,\quad& h_{33}^1&=h_{11}^1\,,\quad&
h_{22}^2&=0\,,\quad& h_{44}^1&=h_{22}^2\,,
\end{align}
for~$\mu=1$, and otherwise,
\begin{align}
h_{14}^1&=h_{23}^1\,\mu\,,\quad& 
h_{33}^1&=\frac{h_{11}^1+(1-\mu)h_{23}^1}{\mu}\,,&\nonumber\\
h_{22}^1&=h_{23}^1\,\mu\,,\quad& h_{22}^2&=0\,,\quad&h_{44}^1&=
h_{23}^1\,\mu\,,
\end{align}
with which we have a candidate symmetrizer. The last part 
of the calculation is to choose the remaining 
parameters so that the candidate symmetrizer is positive 
definite. With the Lorenz gauge~$\mu=1$ the choice 
\begin{align}
h_{11}^1&=1\,,\quad& h_{23}^1&=0\,, \quad& h_{22}^1&=1\,,
\quad& a_{22}^1&=0\,,\quad& h_{22}^1&=0\,, 
\end{align}
does the trick. Other choices work just as well. In the generic 
case the candidate is positive with~$a_{22}^1=0$ 
and~$h_{23}^1<\frac{1}{2+\mu}$. 

Having shown symmetric hyperbolicity we could write down Maximally 
dissipative boundary conditions that render the IBVP well-posed. 
As we will see in the next section, these boundary conditions 
would still not be satisfactory. There are numerical methods, 
called summation by parts methods, reviewed in detail 
elsewhere~\cite{SarTig12}, that can use the conserved energy to 
guarantee stability in numerical approximation. These calculations 
are performed, following~\cite{HilRic10}, in the 
mathematica notebook {\tt Maxwell\_Symmetric.nb} with the 
package {\tt xTensor} for abstract tensor computer 
algebra~\cite{xAct_web}, albeit in a trivial way. 

\subsubsection{Application of the Laplace-Fourier method}

\begin{figure}[pt]
\centerline{\psfig{file=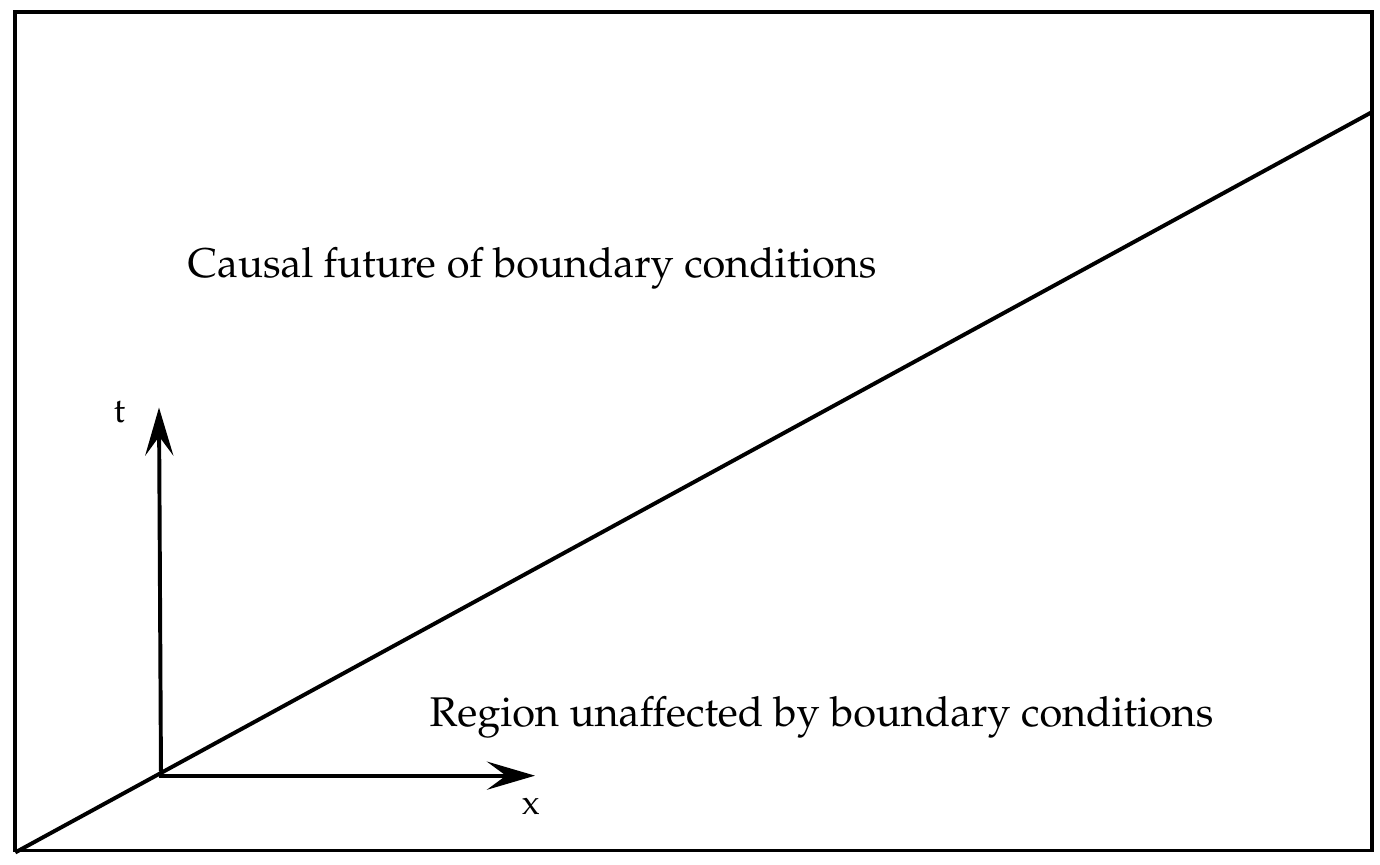,width=9cm}}
\vspace*{8pt}
\caption{Besides being well-posed, the boundary conditions at~$x=0$ 
should be constraint preserving, otherwise everything in their causal 
future will not solve field equations. In this context causal future 
means ``what can be affected by improper boundary conditions''. In 
our formulation~\eqref{eqn:Max_Full} constraint violation propagates 
at the speed of light, but one could equally well construct systems 
in which constraint violation propagates faster. In numerical 
applications error can travel faster than the maximum speed present. 
Care is needed in interpreting the affected region.~\label{fig:BCs}}
\end{figure}

Electromagnetism has a crucial complication that was 
not present in the examples that we considered with the 
IBVP in the last lecture. The presence of constraints 
complicates the analysis of the IBVP. If the formulation 
is symmetric hyperbolic, we can use maximally dissipative 
boundary conditions to guarantee well-posedness, but 
such boundary conditions will in general not be compatible 
with the constraints. In other words, their use will pump 
constraint violation into the domain from the boundary, 
as shown in Fig.~\ref{fig:BCs}, rendering the solution to the IBVP 
unphysical. With this in mind we require three 
properties~\cite{Ste98a,Rin06a,RuiRinSar07} of our boundary 
conditions:
\begin{romanlist}[(ii).]
\item Well-posedness, which, with all that has been 
said already we need not comment on here.
\item Constraint preservation. We are interested in 
computing solutions to the Maxwell equations, so 
the boundary conditions had better respect the 
constraints. In numerical approximation, it is ok 
for the boundary conditions to cause violation of the 
constraints, provided that the violation converges away
with resolution. This is only possible if the underlying
continuum boundary conditions are constraint preserving. 
\item Radiation control. The boundary conditions should
also control the incoming physical radiation in a way 
that is appropriate to the problem at hand. In general relativity, 
we are typically modeling asymptotically flat space–times, 
so this really means {\it no incoming gravitational waves}.
But in electromagnetism we might be interested 
in the response of the field to an incoming wave.
\end{romanlist}
In the context of numerical relativity, we consider the initial 
boundary value problem because computers have only finite memory, 
and so modeling the entire physical domain accurately is difficult.
One possible way to simulate the entire region of interest which is 
currently under investigation is to use~``hyperboloidal'' slices, which 
are spacelike, but which terminate at future null 
infinity~\cite{RinMon13,BerNagZen11,Rin09,Zen08,ZenHus06,CalGunHil05,Fra04}. 
Another would be to use methods called cauchy-characteristic-matching, or 
cauchy-characteristic-extraction, in which a code that solves the field 
equations in the form we've been considering communicates data to another 
code that solves the equations in null coordinates all of the way out to 
null infinity~\cite{Win98,ReiBisPol09,ReiBisPol09a,ReiBisPol12}.
Nevertheless the current standard in numerical relativity is to truncate 
the domain at some large radius. The communication needed in the 
cauchy-characteristic procedure is also expected to require an understanding 
of the initial boundary value problem, as is the use of hyperboloidal slices
if such data are evolved with a formulation that has superluminal speeds.

For the Laplace-Fourier analysis we work in the spatial half 
plane~$x\geq0$. We fix,
\begin{align}
\alpha&=1\,,\quad& \beta^x&=0\,,\quad&\beta^y&=0\,,\quad&
\beta^z&=0\,,
\end{align}
and assume that~$\gamma_{ij}$ is just the identity. The only 
one of these assumptions that is not justified is 
that~$\beta^x=0$~\cite{RuiRinSar07}, which can also be dealt 
with~\cite{HilRui12}, but which makes the algebra much more complicated 
and does not help to illustrate how the method is to be applied. 
Performing the Laplace-Fourier transform, and eliminating~$E_i$ 
and~$Z$ gives a set of second order ODEs
\begin{align}
s^2\,\hat{\Phi}&=\mu\,[\p_x^2-\omega^2]
\hat{\Phi}\,,\nonumber\\
s^2\,\hat{A}_x&=[\p_x^2-\omega^2]\hat{A}_x
+\big(\tfrac{1}{\mu}-1\big)
s\,\p_x\hat{\Phi}\,,\nonumber\\
s^2\,\hat{A}_{\hat{\omega}}&=
[\p_x^2-\omega^2]\hat{A}_{\hat{\omega}}+\big(\tfrac{1}{\mu}-1\big)\,i\,\omega\,
s\,\hat{\Phi}\,,\nonumber\\
s^2\,\hat{A}_{\hat{\nu}}&=[\p_x^2-\omega^2]\hat{A}_{\hat{\nu}}\,,
\end{align}
where the vector~$\hat{A}_i$ has been decomposed according to
\begin{align}
\hat{A}_i&=\hat{x}_i\hat{A}_{\hat{x}}
+\hat{\omega}_i\hat{A}_{\hat{\omega}}
+\hat{\nu}_i\hat{A}_{\hat{\nu}}.
\end{align}
with~$\hat{x}^i$, a unit vector in the x-direction,~$\hat{\omega}^i$
a unit vector in the~$\omega^i$ direction, and~$\hat{\nu}^i$
a unit vector orthogonal to both~$\hat{x}^i$ and~$\hat{\omega}^i$.
Exactly as described in the first lecture, we reduce the system to a 
set of first order ODEs by introducing the pseudo-differential 
reduction variables,
\begin{align}
D\hat{\Phi}&=\p_x\hat{\Phi}/\kappa\,,
\quad&D\hat{A}_x&=\p_x\hat{A}_x/\kappa\,,\quad&
D\hat{A}_{\hat{\omega}}&=\p_x\hat{A}_{\hat{\omega}}/\kappa\,,\quad&
D\hat{A}_{\hat{\nu}}&=\p_x\hat{A}_{\hat{\nu}}/\kappa\,,
\end{align}
where~$\omega=\sqrt{\omega^i\omega_i}$, and we 
define~$\kappa=\sqrt{|s|^2+\omega^2}$. We will also use the 
shorthands
\begin{align}
s'&=\kappa^{-1}s\quad& \omega'&=\kappa^{-1}\omega\,\nonumber\\
\tau_{\pm}&=\pm\kappa\,\sqrt{s'^2+\omega'^2}\quad&
\tau_{\pm}'&=\kappa^{-1}\tau_{\pm}\nonumber\\
\tau_{\pm\mu}&=\pm\kappa\,\sqrt{\tfrac{s'^2}{\mu}+\omega'^2},\quad&
\tau_{\pm\mu}'&=\kappa^{-1}\tau_{\pm\mu}\,.
\end{align}
The first order ODE system is, 
\begin{align}
\p_x\hat{\Phi}&=\kappa\,D\hat{\Phi}\,,\quad&
\p_xD\hat{\Phi}&=-\kappa\,\tau_{+\mu}'\,\tau_{-\mu}'\hat{\Phi}\,,\nonumber\\
\p_x\hat{A}_x&=\kappa\,D\hat{A}_x\,,\quad&
\p_xD\hat{A}_x&=-\kappa\,\tau_{+}'\,\tau_{-}'\,\hat{A}_x
+\kappa\,\big(1-\tfrac{1}{\mu}\big)\,s'\,D\hat{\Phi}\,,\nonumber\\
\p_x\hat{A}_{\hat{\omega}}&=\kappa\,D\hat{A}_{\hat{\omega}}\,,\quad&
\p_xD\hat{A}_{\hat{\omega}}&=-\kappa\,\tau_{+}'\,\tau_{-}'\,\hat{A}_{\hat{\omega}}
+i\,\omega'\kappa\,\big(1-\tfrac{1}{\mu}\big)\,s'\,\hat{\Phi}\,,\nonumber\\
\p_x\hat{A}_{\hat{\nu}}&=\kappa\,D\hat{A}_{\hat{\nu}}\,,\quad&
\p_xD\hat{A}_{\hat{\nu}}&=-\kappa\,\tau_{+}'\,\tau_{-}'\,\hat{A}_{\hat{\nu}}\,,
\end{align}
which is obviously of the form~$\p_x\hat{u}=M\,\hat{u}$. Since in this 
case~$M$ is always diagonalizable, the general~$L_2$ solution to this system 
at the boundary~$x=0$ is formed by a sum over the eigenvectors of~$M$ whose 
eigenvector has negative real part. It is,
\begin{align}
\hat{\Phi}&=\sigma_\Phi\,,\quad\quad&
D\hat{\Phi}&=\tau_{-\mu}'\sigma_\Phi\,,\nonumber\\
\hat{A}_x&=-\frac{\sigma_Z}{\tau_-'}
-\frac{\tau_{-\mu}'\sigma_{\Phi}}{s'}
-\frac{i\,\omega'\,\sigma_{A_{\hat{\omega}}}}{\tau_-'}\,,\quad&
D\hat{A}_x&=-\sigma_Z
-\frac{\tau_{-\mu}'^2\sigma_{\Phi}}{s'}-i\,\omega'\,
\sigma_{A_{\hat{\omega}}}\,,\nonumber\\
\hat{A}_{\hat{\omega}}&=\sigma_{A_{\hat{\omega}}}
-\frac{i\,\omega'\,\sigma_{\Phi}}{s'}\,,\quad&
D\hat{A}_{\hat{\omega}}&=\tau_{-}'\sigma_{A_{\hat{\omega}}}
-\frac{i\,\omega'\,\tau_{-\mu}'\,\sigma_{\Phi}}{s'}\,,\nonumber\\
\hat{A}_{\hat{\nu}}&=\sigma_{A_{\hat{\nu}}}\,,\quad&
D\hat{A}_{\hat{\nu}}&=\tau_{-}'\sigma_{A_{\hat{\nu}}}\,,
\end{align}
with~$\sigma_\Phi\,,\sigma_Z\,,\sigma_{A_{\hat{\omega}}}\,,\sigma_{A_{\hat{\nu}}}$ 
complex constants to be determined by substituting into boundary 
conditions, which are yet to be specified. Notice that I have carefully 
constructed the solution so that it is a sum of a gauge~$\sigma_\Phi$,
constraint violating~$\sigma_Z$ and 
physical~$\sigma_{A_{\hat{\omega}}},\sigma_{A_{\hat{\nu}}}$ part. It has not been 
shown that this works for any constrained Hamiltonian system, but using that 
of~\cite{HilRic13}, I expect that this can be done. In any case for
general relativity, you {\it can} write the general solution like 
this~\cite{HilRui12}. Since in the principal part, the gauge field~$\Phi$ 
and the constraint~$Z$ satisfy wave equations, the obvious choice is something 
like a Sommerfeld condition,
\begin{align}
[\p_t-\sqrt{\mu}\,\p_x]^2\Phi&\hateq \p_tg_\Phi\,,\quad&
[\p_t-\p_x]Z&\hateq \p_tg_Z\,,
\end{align}
on each, which should to absorb outgoing waves without causing 
large reflections. In applications we would of course choose~$g_Z=0$, 
so that the incoming characteristic variable of the constraint 
subsystem vanishes. But here we choose an arbitrary function to show that 
boundary stability can be obtained even with non-trivial data for the 
constraints. With constraint preservation out of the way, we have to think about 
good physical boundary conditions. The electric and magnetic fields are gauge 
invariant, and unambiguously represent the Maxwell field strength, 
so we choose,
\begin{align}
[\p_t-\p_x]\big(\p_tA_A+\p_A\Phi-\p_xA_A+\p_AA_x\big)&=\p_tg_A\,.
\end{align}
I leave it as an exercise for you to convince yourself that this is 
really a boundary condition on some combination of the electric and
magnetic fields, actually $\phi_0$ in the terminology of 
Teukolsky~\cite{Teu73}. Laplace-Fourier transforming the boundary conditions 
and substituting the general solution into them, we can solve for the 
constants~$\sigma_\Phi\,,\sigma_Z\,,\sigma_{A_{\hat{\omega}}}\,,\sigma_{A_{\hat{\nu}}}$
and obtain the solution 
\begin{align}
\hat{\Phi}&=\frac{s'\,\hat{g}_\Phi}
{\big(\,s'+\sqrt{s'^2+\mu\,\omega'^2}\,\big)^2}\,,\quad\quad
D\hat{\Phi}=-\frac{\sqrt{s'^2+\mu\,\omega'^2}\,s'\,\hat{g}_\Phi}
{\sqrt{\mu}\,\big(\,s'+\sqrt{s'^2+\mu\,\omega'^2}\,\big)^2}\,,\nonumber\\
\hat{A}_x&=
\frac{i\,\omega'\,\hat{g}_{\hat{\omega}}}{\big(\,s'+\sqrt{s'^2+\omega'^2}\,\big)^2}
+\frac{\hat{g}_{Z}}{s'+\sqrt{s'^2+\omega'^2}}
+\frac{\sqrt{s'^2+\mu\,\omega'^2}\,\hat{g}_{\Phi}}
{\sqrt{\mu}\,\big(\,s'+\sqrt{s'^2+\mu\,\omega'^2}\,\big)^2}\,,\nonumber\\
D\hat{A}_x&=
-\frac{i\,\omega'\,\sqrt{s'^2+\omega'^2}\,\hat{g}_{\hat{\omega}}}
{\big(\,s'+\sqrt{s'^2+\omega'^2}\,\big)^2}
-\frac{\sqrt{s'^2+\omega'^2}\,\hat{g}_{Z}}{s'+\sqrt{s'^2+\omega'^2}}
-\frac{(s'^2+\mu\,\omega'^2)\,\hat{g}_{\Phi}}
{\mu\,\big(\,s'+\sqrt{s'^2+\mu\,\omega'^2}\,\big)^2}\,,\nonumber\\
\hat{A}_{\hat{\omega}}&=
\frac{\sqrt{s'^2+\omega'^2}\,\hat{g}_{\hat{\omega}}}
{\big(\,s'+\sqrt{s'^2+\omega'^2}\,\big)^2}
-\frac{i\,\omega'\,\hat{g}_{Z}}{\big(\,s'+\sqrt{s'^2+\omega'^2}\,\big)^2}
-\frac{i\,\omega'\,\hat{g}_{\Phi}}
{\big(\,s'+\sqrt{s'^2+\mu\,\omega'^2}\,\big)^2}\,,\nonumber\\
D\hat{A}_{\hat{\omega}}&=
-\frac{(s'^2+\omega'^2)\,\hat{g}_{\hat{\omega}}}
{\big(\,s'+\sqrt{s'^2+\omega'^2}\,\big)^2}
+\frac{i\,\omega'\,\sqrt{s'^2+\omega'^2}\,\hat{g}_{Z}}
{\big(\,s'+\sqrt{s'^2+\omega'^2}\,\big)^2}
+\frac{i\,\omega'\,\sqrt{s'^2+\mu\,\omega'^2}\,\hat{g}_{\Phi}}
{\sqrt{\mu}\,\big(\,s'+\sqrt{s'^2+\mu\,\omega'^2}\,\big)^2}\,,\nonumber\\
\hat{A}_{\hat{\nu}}&=\frac{s'\,\hat{g}_{\hat{\nu}}}
{\big(\,s'+\sqrt{s'^2+\omega'^2}\,\big)^2}\,,\quad\quad
D\hat{A}_{\hat{\nu}}=-\frac{s'\,\sqrt{s'^2+\omega'^2}\,\hat{g}_{\hat{\nu}}}
{\big(\,s'+\sqrt{s'^2+\omega'^2}\,\big)^2}\,,
\end{align}
at the boundary~$x=0$, where now I've abandoned most of the shorthands so 
that you can see how it really looks. All that remains is to show that each 
of the variables is bounded in~$s$ and~$\omega$, which I leave as 
an exercise, but point you towards~\cite{KreWin06} which shows 
the necessary estimates on all of the terms present here. The essential 
point is that we don't have to worry about the numerators in any of the 
fractions, because they are obviously bounded, and since the real part 
of~$s'$ is positive, terms like~$s'+\sqrt{s'^2+\omega'^2}$ appearing in 
the denominators are bounded away from zero. Recalling that by 
construction~$\sigma_Z$ corresponds directly to constraint violation, and 
noting that,
\begin{align}
\sigma_Z=\frac{s'\,\hat{g}_Z}{s'+\sqrt{s'^2+\omega'^2}}\,,
\end{align}
it is clear that the constraint preserving boundary conditions work 
when~$g_Z=0$ is chosen.

These calculations are largely performed in the mathematica 
notebook {\tt Maxwell\_LF.nb} which accompanies the lecture.
With the Lorenz gauge,~$\mu=1$, well-posedness can be 
shown for constraint preserving boundary conditions using 
the energy method~\cite{KreReuSar07,KreReuSar08} with a special 
choice of symmetrizer, or alternatively with the Kreiss-Winicour 
cascade approach~\cite{KreWin06}. To my knowledge this is 
the first time that boundary stability has been demonstrated 
for arbitrary hyperbolic gauge conditions inside the 
family~\eqref{eqn:Gauge}.

\subsection{Summary}

In this lecture we looked at different formulations of electromagnetism 
suitable for free-evolution. We saw that for every strongly hyperbolic 
pure gauge, we could build a formulation which was itself strongly 
hyperbolic. We saw furthermore that system is system is symmetric 
hyperbolic for all of these gauge choices. Working then in the 
high-frequency frozen coefficient approximation, we used the 
Laplace-Fourier method to investigate boundary stability with 
constraint preserving boundary conditions. Some parts of the calculations 
were not very explicitly presented. To understand the ins-and-outs I 
recommend that you study the mathematica notebooks in tandem with the 
lecture notes. They are available at the school's 
website~\cite{NRHEP2_web}. Some of the calculations presented in this 
lecture used the mathematica package {\it xTensor}~\cite{xAct_web} for 
abstract tensor calculations.

\section*{Acknowledgments}

I would like to thank the organizers of the NRHEP2 school for giving 
me the rewarding opportunity to present these lectures. I have benefited 
greatly from discussions with Bernd Br\"ugmann, Ronny Richter, Milton 
Ruiz and Andreas Weyhausen. I am especially grateful to Olivier Sarbach
for carefully reading the lecture notes and offering very helpful 
criticism. I am supported by the DFG grant SFB/Transregio~$7$.

\bibliographystyle{unsrt}

\begin{thebibliography}{10}

\bibitem{BauSha10}
Thomas Baumgarte and Stuart Shapiro.
\newblock {\em Numerical Relativity: Solving {E}instein's Equations on the
  Computer}.
\newblock Cambridge University Press, Cambridge, 2010.

\bibitem{Alc08}
Miguel Alcubierre.
\newblock {\em Introduction to 3+1 Numerical Relativity}.
\newblock Oxford University Press, USA, 2008.

\bibitem{Spe13}
Ulrich Sperhake.
\newblock {Numerical relativity in higher dimensions}.
\newblock ArXiv:1301.3772.
\newblock 2013.

\bibitem{Pfe12}
Harald~P. Pfeiffer.
\newblock {Numerical simulations of compact object binaries}.
\newblock {\em Class.Quant.Grav.}, 29:124004, 2012.

\bibitem{SarTig12}
Olivier Sarbach and Manuel Tiglio.
\newblock Continuum and discrete initial-boundary value problems and einstein's
  field equations.
\newblock {\em Living Reviews in Relativity}, 15(9), 2012.

\bibitem{Hin10}
Ian Hinder.
\newblock {The Current Status of Binary Black Hole Simulations in Numerical
  Relativity}.
\newblock {\em Class.Quant.Grav.}, 27:114004, 2010.

\bibitem{Spe09}
U.~Sperhake.
\newblock {Colliding black holes and gravitational waves}.
\newblock {\em Lect. Notes Phys.}, 769:125--175, 2009.

\bibitem{GraNov09}
Philippe Grandcl{\'e}ment and J{\'e}r{\^o}me Novak.
\newblock Spectral methods for numerical relativity.
\newblock {\em Living Reviews in Relativity}, 12(1), 2009.

\bibitem{Gou10}
Eric Gourgoulhon.
\newblock {An Introduction to the theory of rotating relativistic stars}.
\newblock arXiv:1003.5015.
\newblock 2010.

\bibitem{Gou07}
Eric Gourgoulhon.
\newblock 3+1 formalism and bases of numerical relativity.
\newblock arXiv:gr-qc/0703035.
\newblock 2007.

\bibitem{Pre07a}
Frans Pretorius.
\newblock Binary black hole coalescence.
\newblock In W.~B. Burton, editor, {\em Physics of Relativistic Objects in
  Compact Binaries: From Birth to Coalescence}, volume 359, pages 305--369.
  Springer, Netherlands, 2009.

\bibitem{KreLor89}
Heinz~Otto Kreiss and J.~Lorenz.
\newblock {\em Initial-boundary value problems and the {N}avier-{S}tokes
  equations}.
\newblock Academic Press, New York, 1989.

\bibitem{GusKreOli95}
Bertil Gustafsson, Heinz-Otto Kreiss, and Joseph Oliger.
\newblock {\em Time dependent problems and difference methods}.
\newblock Wiley, New York, 1995.

\bibitem{KnaWalBau02}
A.~M. Knapp, E.~J. Walker, and Thomas~W. Baumgarte.
\newblock Illustrating stability properties of numerical relativity in
  electrodynamics.
\newblock {\em Phys. Rev. D}, 65:064031, 2002.

\bibitem{LinSchKid04}
Lee Lindblom, Mark~A. Scheel, Lawrence~E. Kidder, Harald~P. Pfeiffer, Deirdre
  Shoemaker, and Saul~A. Teukolsky.
\newblock Controlling the growth of constraints in hyperbolic evolution
  systems.
\newblock {\em Phys. Rev. D}, 69:124025, 2004.

\bibitem{Reu04}
Oscar~A. Reula.
\newblock {Strongly hyperbolic systems in general relativity}.
\newblock {\em Journal of Hyperbolic Differential Equations}, 1:251--269, 2004.

\bibitem{GunGar04}
Carsten Gundlach and Jose M. Martin-Garcia.
\newblock Symmetric hyperbolic form of systems of second-order evolution
  equations subject to constraints.
\newblock {\em Phys. Rev. D}, 70:044031, 2004.

\bibitem{GunGar05}
Carsten Gundlach and Jose M. Mart{\'\i}n-Garc{\'\i}a.
\newblock Hyperbolicity of second-order in space systems of evolution
  equations.
\newblock {\em Class. Quantum Grav.}, 23:S387--S404, 2006.

\bibitem{ReuSar05}
Oscar Reula and Olivier Sarbach.
\newblock {A Model problem for the initial-boundary value formulation of
  Einstein's field equations}.
\newblock {\em J.Hyperbol.Diff.Equat.}, 2:397, 2005.

\bibitem{AlcDegSal09}
Miguel Alcubierre, Juan~Carlos Degollado, and Marcelo Salgado.
\newblock {The Einstein-Maxwell system in 3+1 form and initial data for
  multiple charged black holes}.
\newblock {\em Phys.Rev.}, D80:104022, 2009.

\bibitem{MacPfe13}
Harald~P. Pfeiffer and Andrew~I. MacFadyen.
\newblock {Hyperbolicity of Force-Free Electrodynamics}.
\newblock arXiv:1307.7782.
\newblock 2013.

\bibitem{Kre59}
Heinz-Otto Kreiss.
\newblock {\"U}ber matrizen die beschr{\"a}nkte halbgruppen erzeugen.
\newblock {\em Math. Scand.}, 7:71--80, 1959.

\bibitem{MajOsh75}
A.~Majda and S.~Osher.
\newblock Initial-boundary value problems for hyperbolic equations with
  uniformly characteristic boundary.
\newblock {\em Commun. Pure Appl. Math.}, 28:607–675, 1975.

\bibitem{LaxPhi60}
P.~D. Lax and R.~S. Phillips.
\newblock Local boundary conditions for dissipative symmetric linear
  differential operators.
\newblock {\em Commun. Pure Appl. Math.}, 13:427--455, 1960.

\bibitem{Rau85}
J.~Rauch.
\newblock Symmetric positive systems with boundary characteristics of constant
  multiplicity.
\newblock {\em Trans. Am. Math. Soc.}, 291:167, 1985.

\bibitem{Sec96}
Paolo Secchi.
\newblock The initial boundary value problem for linear symmetric hyperbolic
  systems with characteristic boundary of constant multiplicity.
\newblock {\em Differential Integral Equations}, 9:671--700, 1996.

\bibitem{CalSar03}
Gioel Calabrese and Olivier Sarbach.
\newblock Detecting ill posed boundary conditions in general relativity.
\newblock {\em J. Math. Phys}, 44:3888--3889, 2003.

\bibitem{Kre70}
Heinz-Otto Kreiss.
\newblock Initial boundary value problems for hyperbolic systems.
\newblock {\em Comm. Pure Appl. Math.}, 23:277--298, 1970.

\bibitem{Agr72}
M.~S. Agranovich.
\newblock Theorem on matrices depending on parameters and its applications to
  hyperbolic systems.
\newblock {\em Functional Analysis and Its Applications}, 6:85--93, 1972.

\bibitem{Met00}
Guy M\'etivier.
\newblock The block structure condition for symmetric hyperbolic systems.
\newblock {\em Bulletin of the London Mathematical Society}, 32:689--702, 2000.

\bibitem{HilRic13a}
David Hilditch and Ronny Richter.
\newblock {Hyperbolicity of High Order Systems of Evolution Equations}.
\newblock 2013.
\newblock In preparation.

\bibitem{Tay81}
Michael~E. Taylor.
\newblock {\em Pseudodifferential operators / Michael E. Taylor}.
\newblock Princeton University Press, Princeton, N.J. :, 1981.

\bibitem{NagOrtReu04}
G.~Nagy, O.~E. Ortiz, and O.~A. Reula.
\newblock Strongly hyperbolic second order {E}instein's evolution equations.
\newblock {\em Phys. Rev. D}, 70:044012, 2004.

\bibitem{HilRic10}
David Hilditch and Ronny Richter.
\newblock {Hyperbolic formulations of General Relativity with Hamiltonian
  structure}.
\newblock {\em Phys.Rev.}, D86:123017, 2012.

\bibitem{RicHil11}
Ronny Richter and David Hilditch.
\newblock {Hyperbolicity of Hamiltonian formulations in General Relativity}.
\newblock {\em J.Phys.Conf.Ser.}, 314:012102, 2011.

\bibitem{KreWin06}
Heinz-Otto Kreiss and Jeffrey Winicour.
\newblock Problems which are well-posed in a generalized sense with
  applications to the {E}instein equations.
\newblock {\em Class. Quantum Grav.}, 23:S405--S420, 2006.

\bibitem{KreOrtPet10}
H.-O. {Kreiss}, O.~E. {Ortiz}, and N.~A. {Petersson}.
\newblock {Initial-boundary value problems for second order systems of partial
  differential equations}.
\newblock arXiv:1012.1065.
\newblock 2010.

\bibitem{Dir64}
Paul A.~M. Dirac.
\newblock {\em Lectures on quantum mechanics}, volume~2 of {\em Belfer Graduate
  School of Science Monographs Series}.
\newblock Belfer Graduate School of Science, New York, 1964.

\bibitem{Wit13}
Helvi Witek.
\newblock {Lecture Notes: Numerical Relativity in higher dimensional
  spacetimes}.
\newblock {\em IJPMA}, 2013.

\bibitem{BauSha98}
T.~W. Baumgarte and S.~L. Shapiro.
\newblock On the {N}umerical integration of {E}instein's field equations.
\newblock {\em Phys. Rev.}, D59:024007, 1998.

\bibitem{ShiNak95}
M.~Shibata and T.~Nakamura.
\newblock Evolution of three-dimensional gravitational waves: {H}armonic
  slicing case.
\newblock {\em Phys. Rev.}, D52:5428--5444, 1995.

\bibitem{NakOohKoj87}
Takashi Nakamura, {Ken-ichi} Oohara, and Yasufumi Kojima.
\newblock General relativistic collapse to black holes and gravitational waves
  from black holes.
\newblock {\em Prog. Theor. Phys. Suppl.}, 90:1--218, 1987.

\bibitem{Jac99}
John~David Jackson.
\newblock {\em Classical Electrodynamics}.
\newblock Wiley, New York, 3rd edition, 1999.

\bibitem{Oka13}
Hirotada Okawa.
\newblock Initial conditions for numerical relativity -- introduction to
  numerical methods for solving elliptic pdes.
\newblock {\em IJPMA}, 2013.

\bibitem{WeyBerHil11}
Andreas Weyhausen, Sebastiano Bernuzzi, and David Hilditch.
\newblock {Constraint damping for the Z4c formulation of general relativity}.
\newblock {\em Phys. Rev. D}, 85:024038, 2012.

\bibitem{Owe07}
Robert Owen.
\newblock {Constraint Damping in First-Order Evolution Systems for Numerical
  Relativity}.
\newblock {\em Phys.Rev.}, D76:044019, 2007.

\bibitem{GunGarCal05}
Carsten Gundlach, Jose M. Martin-Garcia, G.~Calabrese, and I.~Hinder.
\newblock Constraint damping in the {Z4} formulation and harmonic gauge.
\newblock {\em Class. Quantum Grav.}, 22:3767--3774, 2005.

\bibitem{BroFriHub99}
O.~Brodbeck, S.~Frittelli, P.~H{\"u}bner, and O.~A. Reula.
\newblock Einstein's equations with asymptotically stable constraint
  propagation.
\newblock {\em J. Math. Phys.}, 40:909--923, 1999.

\bibitem{Bro10}
J.~David Brown.
\newblock {Action Principle for the Generalized Harmonic Formulation of General
  Relativity}.
\newblock {\em Phys.Rev.}, D84:084014, 2011.

\bibitem{BonBonPal10}
C.~Bona, C.~Bona-Casas, and C.~Palenzuela.
\newblock {Action principle for Numerical Relativity evolution systems}.
\newblock {\em Phys. Rev.}, D82:124010, 2010.

\bibitem{Bro08}
J.~David Brown.
\newblock {Strongly Hyperbolic Extensions of the ADM Hamiltonian}.
\newblock arXiv:0803.0334.
\newblock 2008.

\bibitem{HilRic13}
David Hilditch and Ronny Richter.
\newblock {Hyperbolicity of Physical Theories with Application to General
  Relativity}.
\newblock arXiv:1303.4783.
\newblock 2013.

\bibitem{xAct_web}
Jose~M. Martin-Garcia.
\newblock x{A}ct: tensor computer algebra.
\newblock {\tt http://www.xact.es/}.

\bibitem{Ste98a}
J.M.Stewart.
\newblock The {C}auchy problem and the initial boundary value problem in
  numerical relativity.
\newblock {\em Class. Quantum Grav.}, 15:2865, 1998.

\bibitem{Rin06a}
Oliver Rinne.
\newblock {Stable radiation-controlling boundary conditions for the generalized
  harmonic Einstein equations}.
\newblock {\em Class. Quant. Grav.}, 23:6275--6300, 2006.

\bibitem{RuiRinSar07}
Milton Ruiz, Oliver Rinne, and Olivier Sarbach.
\newblock Outer boundary conditions for einstein's field equations in harmonic
  coordinates.
\newblock {\em Class. Quant. Grav.}, 24:6349--6378, 2007.

\bibitem{RinMon13}
Oliver Rinne and Vincent Moncrief.
\newblock {Hyperboloidal Einstein-matter evolution and tails for scalar and
  Yang-Mills fields}.
\newblock {\em Class.Quant.Grav.}, 30:095009, 2013.

\bibitem{BerNagZen11}
Sebastiano Bernuzzi, Alessandro Nagar, and Anil Zenginoglu.
\newblock {Binary black hole coalescence in the large-mass-ratio limit: the
  hyperboloidal layer method and waveforms at null infinity}.
\newblock {\em Phys.Rev.}, D84:084026, 2011.

\bibitem{Rin09}
Oliver Rinne.
\newblock {An axisymmetric evolution code for the Einstein equations on
  hyperboloidal slices}.
\newblock {\em Class. Quant. Grav.}, 27:035014, 2010.

\bibitem{Zen08}
Anil Zenginoglu.
\newblock {Hyperboloidal evolution with the Einstein equations}.
\newblock {\em Class. Quant. Grav.}, 25:195025, 2008.

\bibitem{ZenHus06}
Anil Zengino{\u g}lu and Sascha Husa.
\newblock {Hyperboloidal foliations with scri-fixing in spherical symmetry}.
\newblock {\em Class. Quantum Grav.}, 25:19, 2008.

\bibitem{CalGunHil05}
Gioel Calabrese, Carsten Gundlach, and David Hilditch.
\newblock {Asymptotically null slices in numerical relativity: Mathematical
  analysis and spherical wave equation tests}.
\newblock {\em Class.Quant.Grav.}, 23:4829--4846, 2006.

\bibitem{Fra04}
J{\"o}rg Frauendiener.
\newblock Conformal infinity.
\newblock {\em Living Rev. Relativity}, 7(1), 2004.

\bibitem{Win98}
Jeffrey Winicour.
\newblock Characteristic evolution and matching.
\newblock {\em Living Rev. Relativity}, 1(5), 1998.

\bibitem{ReiBisPol09}
C.~Reisswig, N.~T. Bishop, D.~Pollney, and B.~Szilagyi.
\newblock {Unambiguous determination of gravitational waveforms from binary
  black hole mergers}.
\newblock {\em Phys. Rev. Lett.}, 103:221101, 2009.

\bibitem{ReiBisPol09a}
C.~Reisswig, N.T. Bishop, D.~Pollney, and B.~Szilagyi.
\newblock {Characteristic extraction in numerical relativity: binary black hole
  merger waveforms at null infinity}.
\newblock {\em Class.Quant.Grav.}, 27:075014, 2010.

\bibitem{ReiBisPol12}
Christian Reisswig, Nigel~T. Bishop, and Denis Pollney.
\newblock {General relativistic null-cone evolutions with a high-order scheme}.
\newblock {\em Gen.Rel.Grav.}, 45:1069--1094, 2013.

\bibitem{HilRui12}
David Hilditch and Milton Ruiz.
\newblock {The initial boundary value problem of the Z4c formulation of General
  Relativity}.
\newblock 2013.
\newblock In preparation.

\bibitem{Teu73}
Saul~A. Teukolsky.
\newblock Perturbations of a rotating black hole. {I}. fundamental equations
  for gravitational, electromagnetic, and neutrino-field perturbations.
\newblock {\em Astrophys. J.}, 185:635--647, 1973.

\bibitem{KreReuSar07}
H.O. Kreiss, O.~Reula, O.~Sarbach, and J.~Winicour.
\newblock {Well-posed initial-boundary value problem for the harmonic Einstein
  equations using energy estimates}.
\newblock {\em Class.Quant.Grav.}, 24:5973--5984, 2007.

\bibitem{KreReuSar08}
H.-O. Kreiss, O.~Reula, O.~Sarbach, and J.~Winicour.
\newblock {Boundary conditions for coupled quasilinear wave equations with
  application to isolated systems}.
\newblock {\em Commun.Math.Phys.}, 289:1099--1129, 2009.

\bibitem{NRHEP2_web}
{NR/HEP2: Spring School}.
\newblock {\tt http://blackholes.ist.utl.pt/nrhep2/}

\end{thebibliography}


\end{document}